\documentclass[aps,pra,reprint,superscriptaddress,amsmath,amsfonts,amssymb]{revtex4-1}

\usepackage{bm}
\usepackage[pdftex]{graphicx}
\usepackage{color}
\usepackage{MnSymbol}
\usepackage{enumerate}
\usepackage{bbold}
\usepackage{lipsum}
\usepackage{array}
\usepackage{tikz}
\usetikzlibrary{shapes,arrows}
\usetikzlibrary{positioning,arrows}
\usetikzlibrary{decorations.pathmorphing}
\usetikzlibrary{decorations.markings}

\begin{document}
\tikzset{
particle/.style={thick,draw=black, postaction={decorate},decoration={markings,mark=at position .5 with {\arrow[draw]{<}}}},
particle2/.style={thick,draw=black, postaction={decorate},decoration={markings,mark=at position .5 with {\arrow[draw]{<}}}},
gluon/.style={thick,decorate, draw=black,decoration={coil,aspect=0,segment length=3pt,amplitude=1pt}}
 }
\title{Simultaneous continuous measurement of non-commuting observables: \\ quantum state correlations}
\author{Areeya Chantasri$^*$}
\affiliation{Department of Physics and Astronomy, University of Rochester, Rochester, New York 14627, USA}
\affiliation{Center for Coherence and Quantum Optics, University of Rochester, Rochester, New York 14627, USA}
\author{Juan Atalaya$^*$}
\affiliation{Department of Electrical Engineering, University of California, Riverside, California 92521, USA}
\author{Shay Hacohen-Gourgy}
\affiliation{Quantum Nanoelectronics Laboratory, Department of Physics, University  of  California,  Berkeley  CA  94720,  USA}
\affiliation{Center for Quantum Coherent Science, University of California, Berkeley CA 94720, USA}
\author{Leigh S. Martin}
\affiliation{Quantum Nanoelectronics Laboratory, Department of Physics, University  of  California,  Berkeley  CA  94720,  USA}
\affiliation{Center for Quantum Coherent Science, University of California, Berkeley CA 94720, USA}
\author{Irfan Siddiqi}
\affiliation{Quantum Nanoelectronics Laboratory, Department of Physics, University  of  California,  Berkeley  CA  94720,  USA}
\affiliation{Center for Quantum Coherent Science, University of California, Berkeley CA 94720, USA}
\author{Andrew N. Jordan}
\affiliation{Department of Physics and Astronomy, University of Rochester, Rochester, New York 14627, USA}
\affiliation{Center for Coherence and Quantum Optics, University of Rochester, Rochester, New York 14627, USA}
\affiliation{Institute for Quantum Studies, Chapman University, 1 University Drive, Orange, California 92866, USA}
\date{\today}
\begin{abstract}
We consider the temporal correlations of the quantum state of a qubit subject to simultaneous continuous measurement of two non-commuting qubit observables. Such qubit state correlators are defined for an ensemble of qubit trajectories, which has the same fixed initial state and can also be optionally constrained by a fixed final state. We develop a stochastic path integral description for the continuous quantum measurement and use it to calculate the considered correlators. Exact analytic results are possible in the case of ideal measurements of equal strength and are also shown to agree with solutions obtained using the Fokker-Planck equation. For a more general case with decoherence effects and inefficiency, we use a diagrammatic approach to find the correlators perturbatively in the quantum efficiency. We also calculate the state correlators for the quantum trajectories which are extracted from readout signals measured in a transmon qubit experiment, by means of the quantum Bayesian state update. We find an excellent agreement between the correlators based on the experimental data and those obtained from our analytical and numerical results. 
\end{abstract}

\newcommand{\op}[1]{\hat{ #1}}                
\newcommand{\ket}[1]{\lvert#1\rangle}
\newcommand{\bra}[1]{\langle#1\rvert}
\newcommand{\pr}[1]{\ket{#1}\bra{#1}}
\newcommand{\ipr}[2]{\langle #1 | #2 \rangle}
\newcommand{\mean}[1]{\left\langle #1 \right\rangle}
\newcommand{\cw}{\circlearrowright}
\newcommand{\ccw}{\circlearrowleft}
\newcommand{\la}{\langle}
\newcommand{\ra}{\rangle}
\newcommand{\be}{\begin{equation}}
\newcommand{\ee}{\end{equation}}

\maketitle

\section{Introduction}
Continuous weak quantum measurement (CWQM) has attracted much attention in the quantum information science community. This topic has been discussed theoretically in the past few decades~\cite{BookKraus, Caves1986, Menskii1998, Belavkin1992, BookBraginsky, AAV1988, Dalibard1992, BookCarmichael, wisemanmilburn1993, Korotkov1999}, and its experimental study has been motivated by recent developments in superconducting qubit technology~\cite{katz2006coherent, Laloy2010, Vijay2012, hatr13, murch2013observing, Lange2014, camp13}. Such experimental and theoretical efforts have paved the way for interesting applications of CWQMs such as rapid state purification~\cite{Jacobs2003}, quantum feedback~\cite{Vijay2012,Lange2014,wisemanmilburn1993,Ruskov2002,sayr11}, and preparation of entangled states~\cite{ruskov2003entanglement,Riste2013,Roch2014,chantasri2016}. With CWQMs, it is also possible to simultaneously measure non-commuting observables~\cite{Jordan2005,Wei2008,Ruskov2010}, and the  first experimental demonstration of such measurement on a superconducting qubit was realized only last year~\cite{Shay2016noncom}. 

Precise simultaneous measurement of non-commuting observables of a quantum system is forbidden by textbook quantum mechanics. This is so because if one of the measurements were to collapse the system wavefunction to an eigenstate of a measurement operator, a precise or strong measurement of another non-commuting observable would produce an uncertain result. Such measurement incompatibility with strong measurements, however, can be bypassed by using CWQMs. The reason is that the latter are rather imprecise (weak) measurements such that the readout signals have small signal-to-noise ratios (SNRs), and, therefore, the two competing measurements collapse the system state only partially over time, resulting in a quantum state that continuously evolves in a diffusive manner.

Simultaneous continuous measurement of two qubit observables, $\sigma_z$ and $\sigma_\varphi \equiv \sigma_z\cos\varphi + \sigma_x\sin\varphi$, was implemented by weakly coupling a superconducting transmon qubit to two intracavity modes~\cite{Shay2016noncom}. The measurement was effectively a stroboscopic measurement of a fast rotating qubit. In that experiment, the considered angles included $\varphi=0$ (commuting observables) and $\varphi \ne 0$ (non-commuting observables, for example $\varphi = \pi/2$ corresponds to measurement of $\sigma_x$ and $\sigma_z$). The corresponding quantum trajectories were inferred from the measurement readouts and verified using state tomography techniques. In the case of simultaneous measurement of commuting observables~($\varphi=0$), the qubit state evolves, as expected, to either eigenstates of $\sigma_z$ (i.e., measurement induces two state attractors at the Bloch points $z=\pm1$). In contrast, for simultaneous measurement of $\sigma_x$ and $\sigma_z$, the qubit state does not collapse to any eigenstates of such observables. Instead, the quantum trajectories exhibit free diffusion in the Bloch sphere~\cite{Ruskov2010}. The detector readout signals also continuously vary in time in a random fashion and their temporal correlations were analyzed in Ref.~\cite{Atalaya2017cor}. 

The self-correlation function of the detector readouts does not depend on the measurement quantum efficiency, $\eta$, if we consider temporal correlations at sufficiently long times (larger than the reciprocal bandwidth of the detector). The cross-correlation function is not affected by $\eta$ even at vanishing times~\cite{Atalaya2017cor}. This can be understood physically by considering the readout of a non-ideal detector (with $\eta<1$) as the sum of two contributions; namely, one from the readout of an ideal detector ($\eta=1$) and another from a source of uncorrelated noise, whose strength is proportional to $\eta^{-1}-1$. Both contributions then contribute additively to the readout correlators. In contrast, quantum state correlators depend on the quantum efficiency, and their measurement requires detectors with not too small $\eta$, which is nowadays  possible with experimental setups based on superconducting qubits weakly coupled to microwave cavities~\cite{murch2013observing,Shay2016noncom}. In view of this experimental capability, the goal of this paper is to discuss the temporal correlations of the quantum state itself during weak continuous measurements. 

In this work, we consider the statistical properties of the state of a qubit subject to simultaneous continuous measurement of two non-commuting observables. 
We are interested in the temporal correlations of state-dependent quantities over certain sub-ensembles of quantum trajectories. We consider two scenarios (i) quantum state correlators with pre-selection (fixed initial state), and (ii) quantum state correlators with pre- and post-selection (fixed initial and final states). We adopt the calculation technique based on stochastic path integrals~\cite{chantasri2015stochastic} to qubit trajectories in the Bloch sphere. The stochastic path integral description provides a convenient way to explore the statistical properties of sub-ensembles of quantum trajectories by allowing us to impose boundary conditions at the beginning and at the end of each trajectory. We obtain analytical results for the case of ideal measurements of $\sigma_x$ and $\sigma_z$ of equal strength. In the non-ideal case, $\eta<1$, we develop a diagrammatic perturbation theory, similar to the loop expansion in quantum field theory, to calculate the considered sub-ensemble averages in the limit of small $\eta$. Moreover, we calculate the quantum state correlators from the quantum state trajectories monitored in Ref.~\cite{Shay2016noncom} and show that they agree well with our analytical and numerical results. 

We begin our analysis with a review of the quantum trajectory approach to quantum measurement and the stochastic master equation in Sec.~\ref{sec-sde}, and then introduce the stochastic path integral formalism in Sec.~\ref{sec-spi}. The stochastic path integral formalism is then used to compute statistical quantities such as conditional averages and correlation functions, presented specifically for the ``XZ measurement'' (the joint simultaneous measurement of $\sigma_x$ and $\sigma_z$ observables) in Sec.~\ref{sec-corr}. In Sec.~\ref{sec-xzcorr}, we investigate the simplest case, that of an ideal XZ measurement with equal measurement strength, where the quantum state dynamics resembles diffusion on a sphere. The non-ideal case is considered in \ref{sec-nonideal}, using the perturbative expansion of the path integral in terms of diagrams to approximate the statistical averages.
A complementary approach to this physics using the Fokker-Planck is demonstrated in Sec.~\ref{fpe}.  We compare our results with the correlators constructed from the experiment data using superconducting circuits \cite{Shay2016noncom} in Sec.~\ref{sec-exp}, and conclude in Sec.~\ref{conc}.

\section{Background}
\subsection{The stochastic master equation}\label{sec-sde}

We first discuss the stochastic master equation describing the evolution of a qubit subject to simultaneous CWQM of two non-commuting observables. Any qubit observable can be decomposed in terms of the Pauli matrices $\sigma_x,\sigma_y$ and $\sigma_z$. Thus, without loss of generality, we assume that the measured qubit observables are $\sigma_z$ and 
\begin{align}
\label{eq:sigma-varphi}
\sigma_\varphi \equiv\sigma_z\cos\varphi + \sigma_x\sin\varphi.
\end{align}
The cases of $\varphi=0$ and $\varphi = \pi$ correspond to simultaneous measurement of two commuting observables with correlated and anti-correlated measurement results, respectively. For other angles $\varphi$, the two measured observables do not commute. 

Simultaneous continuous measurement of the observables $\sigma_z$ and $\sigma_\varphi$ in the weak coupling regime, given a system state $\rho(t)$, produces the detector readouts~\cite{Atalaya2017cor,wiseman2009quantum,Jacobs2006} $r_z(t)$  and $r_\varphi(t)$ respectively,
\begin{align}
\label{eq:readouts-def}
r_z(t) =& \frac{\Delta r_z}{2}\big({\rm Tr}[\sigma_z\rho(t)] + \sqrt{\tau_z} \xi_z(t)\big),\nonumber \\
r_\varphi(t) =& \frac{\Delta r_\varphi}{2}\big({\rm Tr}[\sigma_\varphi\rho(t)] + \sqrt{\tau_\varphi} \xi_\varphi(t)\big), 
\end{align}
where $\Delta r_{z}$ and $\Delta r_{\varphi}$ are the responses of the $z$- and $\varphi$-detectors, which we can rescale for simplicity so that $\Delta r_{z,\varphi}=2$. The parameters $\tau_z$ and $\tau_\varphi$ are the detector ``characteristic measurement times'', which are defined as the integration time necessary to obtain a SNR of one for the time-averaged measurement readouts~\cite{Korotkov2001}. The quantum efficiency for each measurement channel is defined as $\eta_i = 1/(2\Gamma_i\tau_i)$, where $i=z,\varphi$ and $\Gamma_{z,\varphi}$ are the total (ensemble averaged) dephasing rates for the two measurements. Ideal measurements correspond to the unit quantum efficiency $\eta_i = 1$, whereas, non-ideal measurements correspond to the case of $\eta_i<1$. The noises $\xi_z(t)$ and $\xi_\varphi(t)$ are assumed uncorrelated Gaussian white noises with the two-time correlation functions 
\begin{align}
\label{eq:correlation-xiz-varphi}
\langle \xi_z(t)\xi_z(t')\rangle = \langle \xi_\varphi(t)\xi_\varphi(t')\rangle =\delta(t-t').
\end{align}
The qubit trajectory is described by the stochastic master equation (It\^o interpretation~\cite{Oksendal2003})~\cite{Atalaya2017cor,wiseman2009quantum,Jacobs2006}
\begin{subequations}\label{eq-itosde}
\begin{align}
{\dot x}  =& -(\Gamma_z + \Gamma_\varphi\cos^2 \varphi)x + \frac{\Gamma_\varphi \sin2\varphi}{2}z + (1-x^2)\frac{ \xi_{\varphi}\sin\varphi}{\sqrt{\tau_\varphi}} \nonumber \\
&- xz \left(\frac{\xi_{\varphi}\cos\varphi}{\sqrt{\tau_\varphi}} + \frac{\xi_z}{\sqrt{\tau_z}} \right), \label{eq:EOM-x}\\
{\dot y}  =& -(\Gamma_z + \Gamma_\varphi)y - xy\frac{\xi_{\varphi}\sin\varphi}{\sqrt{\tau_\varphi}} - yz \left(\frac{\xi_{\varphi}\cos\varphi}{\sqrt{\tau_\varphi}} + \frac{\xi_z}{\sqrt{\tau_z}} \right), \label{eq:EOM-y}\\
{\dot z} =& -\Gamma_\varphi\sin^2\varphi\, z + \frac{\Gamma_\varphi\sin2\varphi}{2}x + (1-z^2)\left(\frac{ \xi_{\varphi}\cos\varphi}{\sqrt{\tau_\varphi} }  + \frac{\xi_z}{\sqrt{\tau_z}} \right) \nonumber \\
&- xz \frac{\xi_{\varphi}\sin\varphi}{\sqrt{\tau_\varphi}}, \label{eq:EOM-z}
\end{align}
\end{subequations}
where we have used the Bloch parametrization of the qubit density matrix, $\rho(t) = [\hat\openone + x(t)\sigma_x + y(t)\sigma_z + z(t)\sigma_z]/2$, where $\hat\openone$ is the identity matrix. The evolution equation for the ensemble-averaged qubit trajectory, $\rho_{\rm ens}(t)$, can be obtained from the above It\^o equations by simply dropping the noise terms, which takes the Lindblad form 
\begin{align}
\label{eq:ens-avg-equ}
\dot \rho_{\rm ens} =& \frac{\Gamma_z}{2}\left(\sigma_z\rho_{\rm ens}\sigma_z-\rho_{\rm ens} \right) + \frac{\Gamma_\varphi}{2}\left(\sigma_\varphi\rho_{\rm ens}\sigma_\varphi - \rho_{\rm ens} \right).
\end{align}
We may also include additional terms to Eqs.~\eqref{eq-itosde} and~\eqref{eq:ens-avg-equ}, in the usual way, to account for coherent evolution of the qubit due to a Hamiltonian, $H_{\rm q}$, and environmental decoherence due to a weak coupling of the qubit to unmonitored degrees of freedom~\cite{Atalaya2017cor}.  

\subsection{Stochastic path integral for  XZ measurements}\label{sec-spi}
An alternative description of the stochasticity of the qubit trajectories can be obtained by writing the joint probability density function, ${\cal P}$, of the noises in the measurement readouts and the quantum state trajectory at all times as a (stochastic) path integral. The general treatment of the stochastic path integral formalism for continuous weak measurements is discussed in detail in Refs.~\cite{Chantasri2013,chantasri2015stochastic}. Here, we consider an example for the XZ measurement, where the measured qubit observables are $\sigma_x$ and $\sigma_z$. The more general case of arbitrary angle $\varphi$ is briefly discussed in Sec.~\ref{generalcase}. To construct such stochastic path integral, we need two ingredients. One is the probability densities of the measurement readout noises at each time step, $dt$, and the other element is the deterministic evolution of the quantum state given particular values of the noises. Considering a qubit measurement of duration $T$, which is divided into $N$ steps of duration $dt$, the joint probability density of the noises and the quantum states (at all times) is given by ${\cal P} = \prod_{k=0}^{N-1} P({\bm q}_{k+1} | {\bm q}_k, \xi_{x,k}, \xi_{z,k} ) P(\xi_{x,k})P(\xi_{z,k})$. The terms $\xi_{x,k}$ and $\xi_{z,k}$ are the values of the white noises at the time $t_k = t_0 + k\, d t$, with the probability density $P(\xi_{x,k}) = \sqrt{dt/2\pi}\, \exp\left(-\xi_{x,k}^2\, dt/2\right)$. The vector ${\bm q}_k = \{ x_k, y_k, z_k \}$ denotes the Bloch vector at that time. The transition probability $P({\bm q}_{k+1} | {\bm q}_k, \xi_{x,k}, \xi_{z,k} ) = \delta({\bm q}_{k+1} - {\cal E}[{\bm q}_k, \xi_{x,k},\xi_{z,k}])$ describes the deterministic evolution of the qubit state for given $\xi_{x,k}$ and $\xi_{z,k}$. 

For the XZ measurement, the qubit evolution is dictated by the stochastic master equation~\eqref{eq-itosde} with $\varphi = \pi/2$,
\begin{subequations}\label{eq-XZ-ito}
\begin{eqnarray}
{\dot x}  &=& -\Gamma_{z} x +  (1-x^2)\frac{ \xi_{x}}{\sqrt{\tau_x} } - x z \frac{\xi_{z}}{\sqrt{\tau_z}}, \\
{\dot z} &=& -\Gamma_{x} z +  (1-z^2)\frac{ \xi_{z}}{\sqrt{\tau_z} } -  x z\frac{\xi_{x}}{\sqrt{\tau_x}},  
\end{eqnarray}  
\end{subequations}
where $\Gamma_x$ denotes the dephasing rate due to measurement of $\sigma_x$. Because the Bloch $y$-coordinate does not appear in Eq.~\eqref{eq-XZ-ito}, we can disregard its evolution as long as we are interested in the qubit state evolution on the Bloch $xz$ plane. This is indeed the case if the initial value of such variable is $y(t=t_0)=0$. Then, from Eq.~\eqref{eq:EOM-y}, $y(t)=0$ for all times and we can redefine the state vector ${\bm q}_k = \{ x_k ,z_k \}$. The form of ${\cal E}[{\bm q}_k, \xi_{z,k}, \xi_{x,k}]$ is obtained by writing Eq.~\eqref{eq-XZ-ito} in an explicit time-discretized form, e.g., ${\dot x} = (x_{k+1}-x_k)/dt = f_x(x_k, z_k, \xi_{x,k}, \xi_{z,k})$.

Following the outline presented in Refs.~\cite{Chantasri2013,chantasri2015stochastic}, we write the delta functions for the deterministic evolution as Fourier integrals to express the joint probability density of the noises and qubit state as ${\cal P}\left(\{ {\bm q}_k \}_1^N, \{ \xi_{x,k} \}_0^{N-1}, \{ \xi_{z,k}\}_0^{N-1}|{\bm q}_0\right) \propto \int\!\!{\cal D}{\bm  p}\, e^{-{\cal S}}$. Note that the latter depends on the initial qubit state, ${\bm q}_0$, at the time $t_0$. We refer to the exponent, ${\cal S}$, as the stochastic action and $\int \!\! {\cal D} {\bm p} \equiv \int_{-i \infty}^{i\infty}  \cdots \int_{-i \infty}^{i\infty} \! \prod_{k=0}^{N-1} (2\pi)^{-2}dp_{x,k} dp_{z,k}$. Here, ${\bm p}_k = \{ p_{x,k}, p_{z,k} \}$, introduced by the Fourier representation of the delta functions, are considered as auxiliary integration variables which are regarded as pure imaginary (so that $\mathcal{S}$ is real). Hereon, our notation will be in the time-continuous form, i.e., $p_x(t) = \lim_{dt \rightarrow 0} \{ p_{x,k} \}$, for simplicity and we will use the time-discrete form whenever necessary. The stochastic action ${\cal S}$ for the XZ measurement of a qubit is given by
\begin{align}\label{eq-actionXZ1}
{\cal S} = \int_{t_0}^T \!\!dt \left\{p_x(t) \dot{x}(t) + p_z(t) \dot{z}(t) - {\cal H} \right\},
\end{align}
where ${\cal H} = - \Gamma_z p_x(t) x(t)  - \Gamma_x p_z(t) z(t) +  {\cal H}_x  + {\cal H}_z$ given that
\begin{subequations}\label{eq-actionXZ2}
\begin{align}
 {\cal H}_x =&\frac{1}{\sqrt{\tau_x}} \left[(1-x^2) \xi_x p_x - x z \xi_x p_z \right] - \frac{\xi_x^2}{2},\\
 {\cal H}_z =&\frac{1}{\sqrt{\tau_z}}\left[(1-z^2) \xi_z p_z  - x z \xi_z p_x \right] - \frac{\xi_z^2}{2}, 
\end{align}
\end{subequations}
where the time argument of the noise and qubit variables are omitted for simplicity.

The joint probability density, ${\cal P}$, mentioned above can be used to compute statistical averages of state-dependent quantities, ${\cal A}[{\bm q}(t)]$, which may depend on ${\bm q}(t)$ at various times; e.g., ${\cal A}[{\bm q}(t)] = z(t_1)z(t_2)$ or $x(t_1)z(t_2)$ where $t_1,t_2$ are some intermediate times between $t_0$ and $T$. In particular, the {\it conditional averages} with fixed boundary conditions, ${\bm q}(t_0) = {\bm q}_0 \equiv {\bm q}_{\rm in}$ and ${\bm q}(T) ={\bm q}_N \equiv {\bm q}_{\rm f}$, of the quantum trajectories can be calculated as follows
\begin{align}
\label{eq-generalcondave}
&_{{\bm q}_{\rm f}} \la {\cal A}[{\bm q}(t)] \ra_{{\bm q}_{\rm in}} =\nonumber \\
&\hspace{1cm}\frac{\int_{{\bm q}(t_0)={\bm q}_{\rm in}}^{{\bm q}(T)={\bm q}_{\rm f}}\! {\cal D}{\bm q}\int\!\!{\cal D}\xi\; {\cal A}[{\bm q}(t)] {\cal P}[{\bm q}(t),\xi(t)|{\bm q}_{\rm in}]}{P({\bm q}_{\rm f},T|{\bm q}_{\rm in},t_0)},
\end{align}
where the denominator 
\begin{align}
\label{eq:cond_prob}
P({\bm q}_{\rm f},T|{\bm q}_{\rm in},t_0) = \int_{{\bm q}(t_0) = {\bm q}_{\rm in}}^{{\bm q}(T)={\bm q}_{\rm f}}\!\! {\cal D}{\bm q}\int\!\!{\cal D}\xi\;{\cal P}[{\bm q}(t),\xi(t)|{\bm q}_{\rm in}],
\end{align}
is the conditional probability density to obtain the final state ${\bm q}_{\rm f}$ at time $T$ given the initial state, ${\bm q}_{\rm in}$, at the time $t_0$. We use the notation $\int_{{\bm q}(0) = {\bm q}_{\rm in}}^{{\bm q}(T) = {\bm q}_{\rm f}} \! {\cal D} {\bm q} \equiv \int \cdots \int \prod_{k=1}^{N-1} dx_k dz_k$, which  implies integration over the intermediate qubit states except over the initial and final states, ${\bm q}_0$ and ${\bm q}_N $, respectively. Similarly, $\int \!\! {\cal D} {\xi} \equiv \int \cdots \int \prod_{k=0}^{N-1} d\xi_{x,k}d\xi_{z,k}$. 

\section{State correlations of jointly measured, non-commuting observables}
\label{sec-corr}
In this section we show how to calculate conditional averages with pre- and/or post-selection using the stochastic path integral formalism. We consider conditional averages of the form, e.g., $_{{\bm q}_{\rm f}}\langle z(t_1)x(t_2)\rangle_{{\bm q}_{\rm in}}$. The latter represents the two-time quantum state correlator of the qubit $z$-coordinate at time $t_1$ and the $x$-coordinate at time $t_2$, and the average is over quantum trajectories with initial state ${\bm q}(t_0) = {\bm q}_{\rm in}$ and final state ${\bm q}(T) = {\bm q}_{\rm f}$. Conditional averages with only pre-selection can be obtained by taking the limit $T\to \infty$. We first consider the case which is simple to treat analytically; namely, ideal XZ measurement with detectors of equal measurement strengths ($\Gamma_x=\Gamma_z$). Then, we discuss a diagrammatic perturbation theory for non-ideal measurements. 

\subsection{Conditional averages for ideal XZ measurement}\label{sec-xzcorr}
We will assume that the initial state of the qubit at the time $t=0$ corresponds to some state on the Bloch $xz$ great circle: $x^2+z^2=1$; i.e., $y(0)=0$. From Eq.~\eqref{eq:EOM-y}, we notice that the $y$-coordinate will remain zero during the XZ measurement. Moreover, ideal measurements are characterized by the fact that pure states remain pure during the measurement, even in the case of simultaneous weak measurement of non-commuting observables~\cite{Ruskov2010}. Thus, for the considered ideal XZ measurement, the quantum trajectories can be parametrized by only the polar coordinate, $\theta(t)$,
\begin{align}
\label{eq:polar-coord}
x(t) = \sin\theta(t), \;\;\; y(t) = 0\;\; {\rm and} \;\; z(t)=\cos\theta(t). 
\end{align}
From Eqs.~\eqref{eq:EOM-x} and~\eqref{eq:EOM-z}, we obtain the following equation for the polar coordinate, $\theta(t)$, in the It\^o interpretation
\begin{align}
{\dot \theta}(t) = &  \left(\Gamma_x-\Gamma_z\right)\sin\theta(t) \cos\theta(t) \nonumber \\
& -\sin \theta(t) \frac{\xi_z(t)}{\sqrt{\tau_z}}+ \cos \theta(t) \frac{\xi_x(t)}{\sqrt{\tau_x}}, 
\end{align}
where $\tau_x=1/2\Gamma_x$ and $\tau_z=1/2\Gamma_z$. In particular, for the case of interest of detectors of equal measurement strengths, the above equation reduces to ($\Gamma_x=\Gamma_z=\Gamma_{\rm m}$) 
\be\label{eq:theta-equ}
{\dot \theta}(t)=  \frac{\xi_\theta(t)}{\sqrt{\tau_{\rm m}}},
\ee
where $\tau_{\rm m}=\tau_x=\tau_z$ is the measurement time of both measurement channels, and $\xi_\theta(t) \equiv \cos\theta(t) \xi_x(t) - \sin\theta(t) \xi_z(t)$ can be regarded as a single Gaussian noise term with two-time correlation function: $\langle \xi_\theta(t)\xi_\theta(t')\rangle = \delta(t-t')$. Equation~\eqref{eq:theta-equ} describes free diffusion of the qubit state on the $xz$ great circle. 

The joint probability density of the qubit state, parametrized by the angle $\theta(t)$, and the effective noise, $\xi_\theta(t)$, can be obtained by following the steps discussed in section~\ref{sec-spi}. We find
\begin{align}
\label{eq:calP_theta_xitheta}
\mathcal{P}[\theta(t),\xi_\theta(t)] \propto \int \mathcal{D}p_\theta\,\exp\big(-\mathcal{S}[p_\theta(t),\theta(t),\xi_\theta(t)]\big),
\end{align}
where the exponent is equal to (assuming that the initial time is $t_0=0$)
\begin{align}
\mathcal{S}= \int_{t_0=0}^T dt\,\left\{ip_\theta(t) \big[\dot\theta(t) - \tau_{\rm m}^{-1/2}\xi_\theta(t)\big] + \frac{\xi_\theta^2(t)}{2}\right\},
\end{align}
and the auxiliary integration variables, $p_\theta(t)$, are real. From Eq.~\eqref{eq-generalcondave}, we notice that to calculate conditional averages of state-dependent quantities only, $\mathcal{A}[{\bm q}(t)]$, it is convenient to integrate out the noise $\xi_{\theta}(t)$ and get
\begin{align}
\label{eq:Ptheta-def}
\mathcal{P}[\theta(t)]& \propto \int{\cal D}\xi_\theta\, \mathcal{P}[\theta(t),\xi_\theta(t)],\\
\label{eq:calP_theta}
&\propto \exp\left(-\frac{\tau_{\rm m}}{2}\int_0^Tdt\, \dot\theta^2(t)\right),
\end{align}
which is the probability density functional for each realization of $\theta(t)$, omitting a trivial proportional constant. In Eq.~\eqref{eq:calP_theta}, the angle coordinate should be treated as a coordinate on the real axis; i.e., $\theta\in(-\infty,\infty)$. Then, the conditional average, Eq.~\eqref{eq-generalcondave}, for XZ measurements with detectors of equal measurement strength, can be written as 
\begin{align}
\label{eq:conditional_Atheta}
&_{{\bm q}_{\rm f}} \la {\cal A}[{\bm q}(\theta)] \ra_{{\bm q}_{\rm in}} = \frac{\sum_{n\in \mathbb{Z}}\int_{\theta(0)=\theta_{\rm in}}^{\theta(T)=\theta_{\rm f}+2\pi n}\! {\cal D}\theta\; {\cal A}[{\bm q}(\theta)]{\cal P}[\theta(t)]}{P(\theta_{\rm f},T|\theta_{\rm in},0)},
\end{align}
where the initial state is ${\bm q}_{\rm in} = (\sin\theta_{\rm in},0,\cos\theta_{\rm in})$ and the final state is ${\bm q}_{\rm f} = (\sin\theta_{\rm f},0,\cos\theta_{\rm f})$; such states are parametrized by the angles $\theta_{\rm in},\theta_{\rm f} \in [0,2\pi)$. The sum in Eq.~\eqref{eq:conditional_Atheta} is over all angles, $\theta(T)$, corresponding to the same physical state ${\bm q}_{\rm f}$. The denominator in Eq.~\eqref{eq:conditional_Atheta} is the transition probability,
\begin{align}
\label{eq:cond_prob_theta}
P(\theta_{\rm f},T|\theta_{\rm in},0) = \sum_{n\in \mathbb{Z}}\int_{\theta(0)=\theta_{\rm in}}^{\theta(T)=\theta_{\rm f}+2\pi n}\! {\cal D}\theta\; {\cal P}[\theta(t)].
\end{align}
%

We are interested in calculating the quantum state correlators with pre- and post-selection
\begin{align}
\label{eq:Czz-cond}
C_{zz}(t_1,t_2|{\bm q}_{\rm f},T;{\bm q}_{\rm in},0) \equiv&\;{}_{{\bm q}_{\rm f}}\langle z(t_1)z(t_2)\rangle_{{\bm q}_{\rm in}}, \\
C_{zx}(t_1,t_2|{\bm q}_{\rm f},T;{\bm q}_{\rm in},0) \equiv&\;{}_{{\bm q}_{\rm f}}\langle z(t_1)x(t_2)\rangle_{{\bm q}_{\rm in}}. \label{eq:Czx-cond}
\end{align} 
Because of the rotational symmetry, the correlator $C_{xx}(t_1,t_2|{\bm q}_{\rm f},T;{\bm q}_{\rm in},0)$ can be obtained from the result for $C_{zz}$ by changing $\theta_{\rm in}\to\theta_{\rm in} -\pi/2$ and $\theta_{\rm f}\to\theta_{\rm f}-\pi/2$. 

In order to compute the conditional correlators, it is convenient to introduce 
\begin{align}
\label{eq:As1s2}
{\cal A}_{s_1s_2}[{\bm q}(\theta)]\equiv\exp\left[i\int_0^T dt\, J_{s_1s_2}(t)\theta(t)\right]
\end{align}
where $J_{s_1s_2}(t) = s_1\delta(t-t_1) + s_2\delta(t-t_2)$ can be regarded as a source field, see below, and $s_1,s_2=\pm1$. Then, the state correlators of interest can be expressed as
\begin{align}
\label{eq:Czz-v2}
C_{zz}(t_1,t_2|{\bm q}_{\rm f},T;{\bm q}_{\rm in},0) &= \frac{1}{4}\sum_{s_1,s_2=\pm1}\,{}_{{\bm q}_{\rm f}}\langle {\cal A}_{s_1s_2}[{\bm q}(\theta)]\rangle_{{\bm q}_{\rm in}},  \\
C_{xz}(t_1,t_2|{\bm q}_{\rm f},T;{\bm q}_{\rm in},0) &= \frac{1}{4i}\sum_{s_1,s_2=\pm1}s_2\,{}_{{\bm q}_{\rm f}}\langle {\cal A}_{s_1s_2}[{\bm q}(\theta)]\rangle_{{\bm q}_{\rm in}}.\label{eq:Cxz-v2}
\end{align}
In Eq.~\eqref{eq:Cxz-v2}, the coefficient $s_2$ in each term of the sum arises because we are interested in the correlation of the $x$-coordinate at the time $t_2$, $x(t_2) =\sum_{s_2=\pm1}s_2\exp[is_2\theta(t_2)]/2i$, and the $z$-coordinate at the time $t_1$, $z(t_1) = \sum_{s_1=\pm1}\exp[is_1\theta(t_1)]/2$. 

The calculation of each term in the sums of Eqs.~\eqref{eq:Czz-v2}--\eqref{eq:Cxz-v2} requires the calculation of the path integral $\int {\cal D}\theta\; {\cal P}[\theta(t)]{\cal A}_{s_1s_2}[{\bm q}(\theta)]$, see also Eq.~\eqref{eq:conditional_Atheta}. Such a Gaussian path integral can be straightforwardly calculated~\cite{BookFeyHib}. We find 
\begin{align}
\label{eq:path-integral}
&\int_{\theta(0)=\theta_0}^{\theta(T)=\theta_T} {\cal D}\theta\; {\cal P}[\theta(t)]{\cal A}_{s_1s_2}[{\bm q}(\theta)] = \left(\frac{\tau_{\rm m}}{2\pi T}\right)^{1/2}\nonumber \\
&\hspace{1cm}\times \exp\left\{-\int_0^Tdt\,\left[\frac{\tau_{\rm m}}{2}{\dot{\bar{\theta}}}(t)^2 - iJ_{s_1s_2}(t)\bar\theta(t)\right]\right\}, 
\end{align}
where $\bar\theta(t)$ satisfies the (saddle-point) equation 
\begin{align}
\label{eq:theta_bar}
\ddot{\bar \theta} = -\frac{i}{\tau_{\rm m}} J_{s_1s_2}(t),
\end{align}
with the boundary conditions $\bar\theta(0)=\theta_0$ and $\bar\theta(T)=\theta_T$. The solution of Eq.~\eqref{eq:theta_bar} can be written in terms of the corresponding Green's function, $G(t,t')$, which satisfies the equation $\partial_t^2 G(t,t') = \delta(t-t')$ with homogeneous boundary conditions: $G(0,t')=G(T,t')=0$,  
\begin{align}
\label{eq:sol-theta-bar}
\bar\theta(t) = -\frac{i}{\tau_{\rm m}}[s_1G(t,t_1) + s_2G(t,t_2)] + \frac{\theta_T - \theta_0}{T}t + \theta_0.
\end{align}
The Green's function reads explicitly as
\begin{align}\label{eq:Green-func}
G(t,t') = (t-t')\Theta(t-t') - (1-t'/T)t,
\end{align}
where $\Theta(\cdot)$ is the Heaviside step function. Note that the Green's function is symmetric, $G(t,t')=G(t',t)$, and $G(t,t)=-t(1-t/T)$. By inserting Eq.~\eqref{eq:sol-theta-bar} into Eq.~\eqref{eq:path-integral}, we find that the sought path integral is given by
\begin{align}
\label{eq:result-PI}
&\int_{\theta(0)=\theta_0}^{\theta(T)=\theta_T} {\cal D}\theta\; {\cal P}[\theta(t)]{\cal A}_{s_1s_2}[{\bm q}(\theta)] =\sqrt{\frac{\tau_{\rm m}}{2\pi T}}F_{s_1s_2}(t_1,t_2)\times\nonumber \\
& \exp\left\{-\frac{\tau_{\rm m}(\theta_{T} - \theta_0)^2}{2T} + i\frac{\theta_T-\theta_0}{T}(s_1t_1 + s_2t_2) + i(s_1+s_2)\theta_0\right\},
\end{align}
where $F_{s_1s_2}(t_1,t_2)\equiv \exp\left\{\sum_{a,b=1,2}s_as_bG(t_a,t_b)/2\tau_{\rm m}\right\}$ is a coefficient independent of $\theta_0$ and $\theta_T$. Using the result~\eqref{eq:result-PI} in Eq.~\eqref{eq:conditional_Atheta}, we obtain 
\begin{align}
\label{eq:result-cond_avg-Atheta}
&_{{\bm q}_{\rm f}} \la {\cal A}_{s_1s_2}[{\bm q}(\theta)] \ra_{{\bm q}_{\rm in}} = F_{s_1s_2}(t_1,t_2)e^{i\theta_{\rm in}(s_1+s_2)} \nonumber \\
&\hspace{1.25cm}\times\frac{\sum_{n\in\mathbb{Z}}e^{-(\Delta\theta + 2\pi n)^2\tau_{\rm m}/2T + i(\Delta \theta +2\pi n)(s_1t_1 + s_2t_2)/T}}{\sum_{n\in\mathbb{Z}}e^{-(\Delta\theta + 2\pi n)^2\tau_{\rm m}/2T}},
\end{align}
where $\Delta \theta=\theta_{\rm f} - \theta_{\rm in}$. From Eqs.~\eqref{eq:result-cond_avg-Atheta} and~\eqref{eq:Czz-v2}--\eqref{eq:Cxz-v2}, the correlators of interest can be found.

The method discussed above to calculate two-time quantum state correlators can be easily generalized to calculate {\it any} $n$-point correlation function of $x$ or $z$. In such cases, we would need to introduce ${\cal A}_{s_1...s_n}$ as a generalization of Eq.~\eqref{eq:As1s2} with a source field $J_{s_1...s_n}(t) = \sum_{j=1}^n s_j \delta(t-t_j)$ and $s_j=\pm1$. The correlators of interest can be written as, for instance, $_{{\bm q}_{\rm f}}\langle x(t_1)z(t_2)x(t_3)\rangle_{{\bm q}_{\rm in}} = \sum_{{s_1} {s_2} {s_3}} {s_1} {s_3}\, {}_{{\bm q}_{\rm f}}\langle {\cal A}_{s_1s_2s_3}\rangle_{{\bm q}_{\rm in}}/2(2i)^2$. Each term in the latter sum is evaluated following a procedure similar to the calculation of ${}_{{\bm q}_{\rm f}}\langle{\cal A}_{s_1s_2}\rangle_{{\bm q}_{\rm in}}$. 

\begin{figure}[tb!]
\centering
\begin{tabular}{cc}
\includegraphics[width=\linewidth, trim = 1.5cm 0.5cm 2cm 0cm,clip=true]{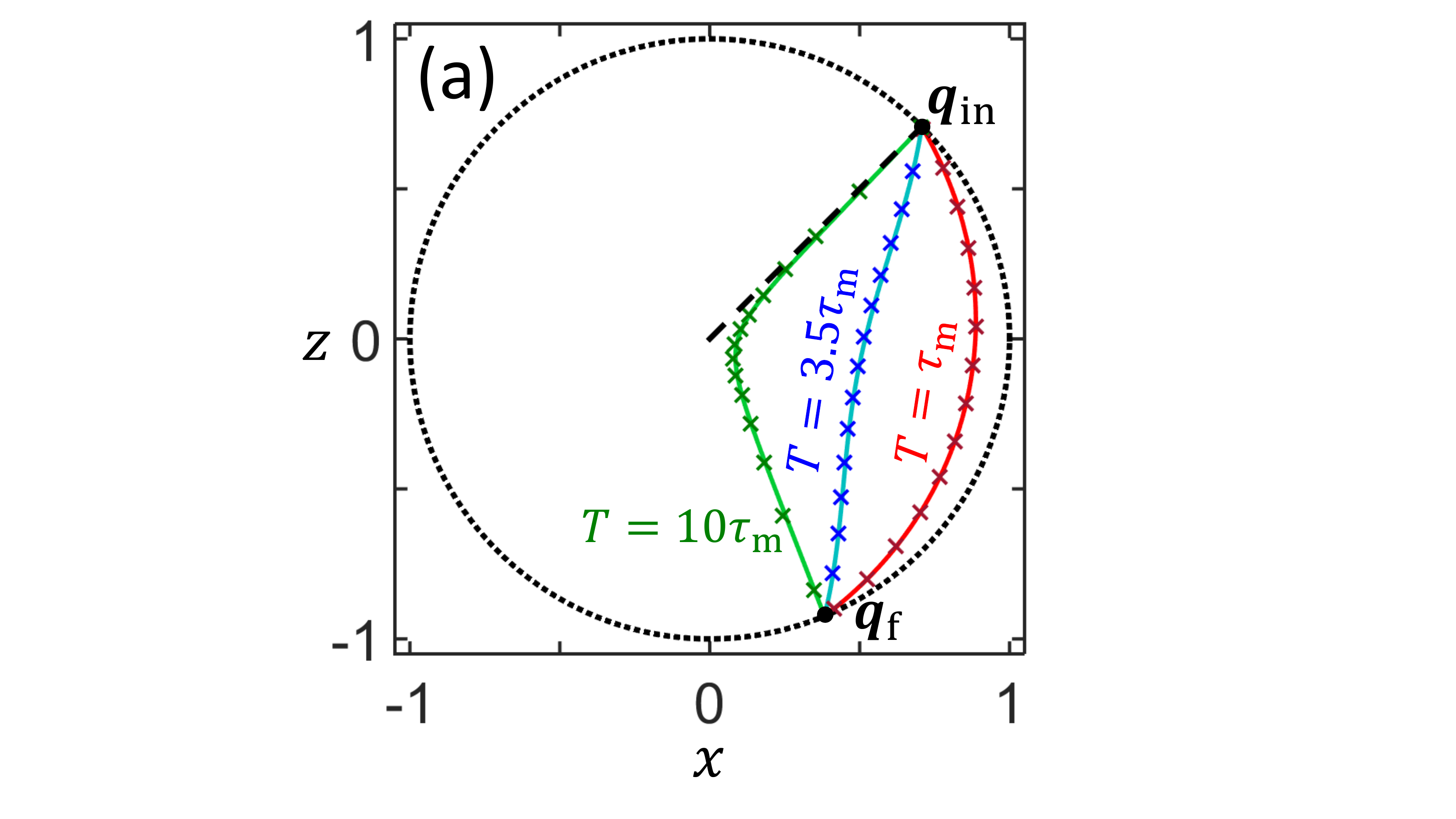}\\
\includegraphics[width=\linewidth, trim = 0.5cm 0.cm 2.5cm 2.0cm,clip=true]{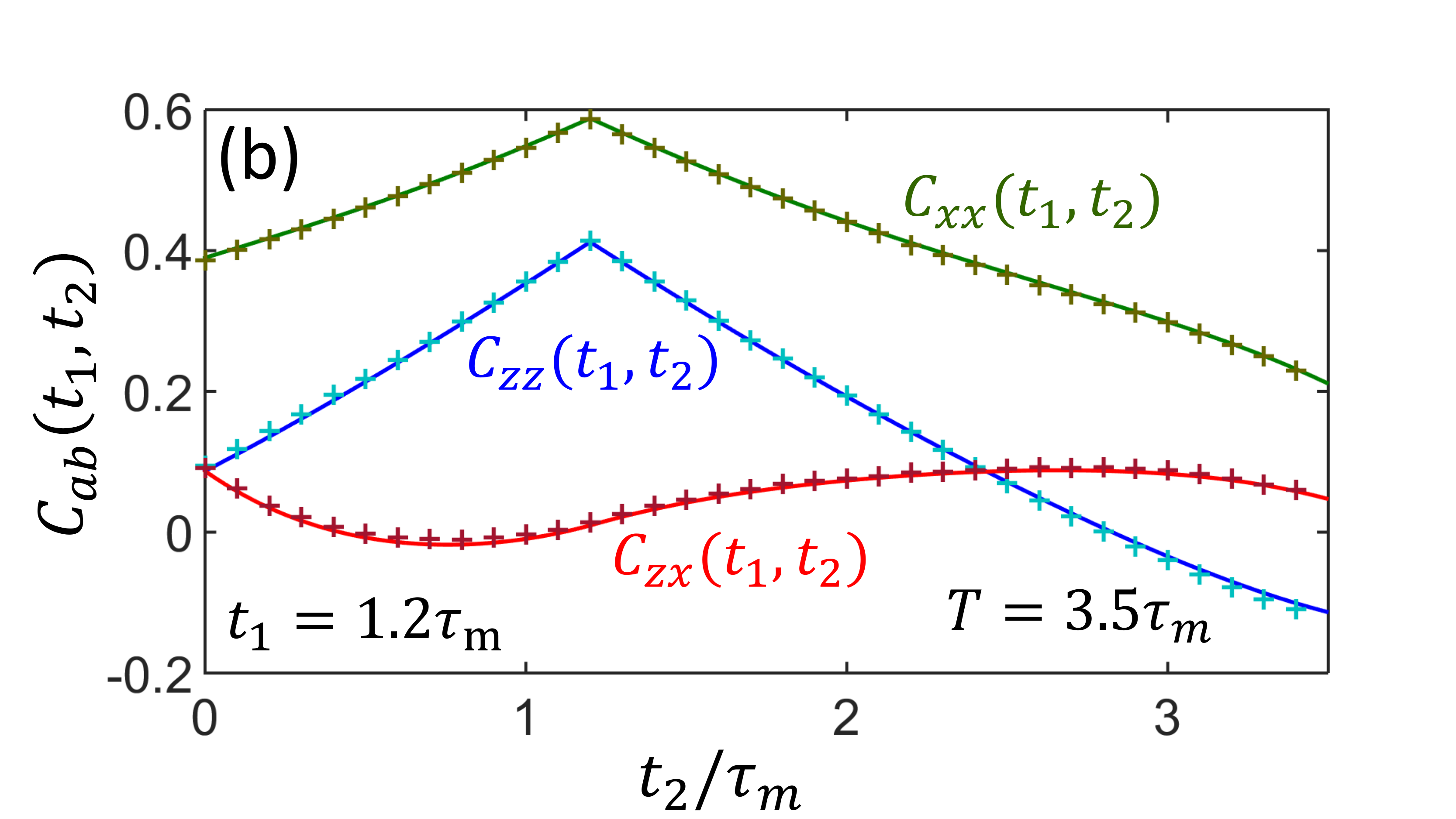}\\
\end{tabular}
\caption{Conditional averages for ideal XZ measurement. Panel (a) shows the qubit state averaged over a sub-ensemble of quantum trajectories, which begin at the initial pure state, ${\bm q}_{\rm in}=\{\cos(\pi/4),0,\sin(\pi/4)\}$ at time $t=0$ and end at the final pure state ${\bm q}_{\rm f}=\{\sin(7\pi/8),0,\cos(7\pi/8)\}$ at time $t=T$. We show the sub-ensemble average trajectories for three different periods $T=\tau_{\rm m}$, $3.5\tau_{\rm m}$ and $10\tau_{\rm m}$. Solid lines depict our analytical results, Eq.~\eqref{eq:sub-ensemble-avg-state}, and the crosses depict Monte Carlo simulation results. The dashed line, connecting the initial state and the fully mixed state, shows the conventional exponential decay of the (total) ensemble average state due to $XZ$ measurement. In panel (b), solid lines show our analytical results for the quantum state correlators, Eqs.~\eqref{eq:Czz-v2}--\eqref{eq:Cxz-v2} and~\eqref{eq:result-cond_avg-Atheta}. The boundary conditions on the quantum trajectories are the same as in panel (a). The crosses represent Monte Carlo simulation results.} 
\label{fig-1}
\end{figure}

\subsubsection{Conditional quantum correlators without post-selection} 

Thus far we have discussed quantum state correlators over sub-ensemble of quantum trajectories with pre- and post-selected states. Now, we consider quantum state correlators without post-selection. Such correlators can be obtained by taking the limit $T \to \infty$ on the previous results. Specifically, 
\begin{align}
\langle {\cal A}_{s_1s_2}[{\bm q}(\theta)]\rangle_{{\bm q}_{\rm in}} &= \lim_{T\to \infty} {}_{{\bm q}_{\rm f}}\langle {\cal A}_{s_1s_2}[{\bm q}(\theta)]\rangle_{{\bm q}_{\rm in}} \\
&=  e^{-(s_1^2t_1 + s_2^2t_2+2s_1s_2t_{\rm min})/2\tau_{\rm m}+ i\theta_{\rm in}(s_1+s_2)},\label{eq:A-preselection}
\end{align}
where $t_{\rm min} = \min\{t_1,t_2\}$. We obtain the result~\eqref{eq:A-preselection} by first applying Poisson's resummation formula to the numerator and denominator in Eq.~\eqref{eq:result-cond_avg-Atheta}; for instance, $\sum_{n\in \mathbb{Z}}\exp[-(\Delta \theta+2\pi n)^2\tau_{\rm m}/2T]=\sqrt{T/2\pi\tau_{\rm m}}\sum_{n\in \mathbb{Z}}\exp[-n^2T/2\tau_{\rm m}+\Delta \theta ni]$. Then, in the limit $T\to \infty$, only the $n=0$ term in the latter sum contributes. The quantum state correlators without post-selection are given by
the following relatively simple closed form expressions
\begin{subequations}
\begin{align}
\la z(t_1) z(t_2) \ra_{{\bm q}_{\rm in}} =\;&  e^{-\frac{t_1+t_2}{2\tau_{\rm m}}} \left\{ \cos^2 \theta_{{\rm in}} \cosh(t_{\rm min}/\tau_{\rm m})  \right.  \nonumber \\ 
& \left. +\sin^2 \theta_{\rm in}  \sinh(t_{\rm min}/\tau_{\rm m})   \right\},\\
\la z(t_1)x(t_2) \ra_{{\bm q}_{\rm in}} =\;& e^{-\frac{t_1+t_2}{2 \tau_{\rm m}}} e^{- \frac{t_{\rm min}}{\tau_{\rm m}}} \sin(2\theta_{\rm in})/2.
\end{align}
\end{subequations}
Notice that the sign of the cross-correlation is determined entirely by the initial angle.  It simply indicates whether the $(x, z)$ coordinates start out as of the same or different sign to give either positive correlation $(+, +)$, $(-,-)$ or negative correlation $(+, -)$, $(-,+)$. 

\subsubsection{Qubit state averaged over sub-ensemble with fixed initial and final states} 

We know the general fact that measurement induces exponential decay of the ensemble average qubit state. In particular, for the considered ideal XZ measurement, the Bloch state decays as $x_{\rm ens}(t)=x_{\rm in}\exp(-\Gamma_{\rm m}t)$, $y_{\rm ens}(t)=y_{\rm in}\exp(-2\Gamma_{\rm m}t)$ and $z_{\rm ens}(t)=z_{\rm in}\exp(-\Gamma_{\rm m}t)$, see Eq.~\eqref{eq:ens-avg-equ}. An interesting question to discuss is how the sub-ensemble average state evolves from a fixed initial state, ${\bm q}_{\rm in}$, at time $t_0$ to a fixed final state ${\bm q}_{\rm f}$ at time $T$.  

To answer the above question, we need to calculate ${\bm q}_{\rm sub-ens-avg}(t)={}_{{\bm q}_{\rm f}}\langle {\bm q}(t)\rangle_{{\bm q}_{\rm in}}$. For the considered ideal XZ measurement, the components ${}_{{\bm q}_{\rm f}}\langle z(t)\rangle_{{\bm q}_{\rm in}}$ and ${}_{{\bm q}_{\rm f}}\langle x(t)\rangle_{{\bm q}_{\rm in}}$ can be calculated from the real and imaginary parts of ${}_{{\bm q}_{\rm f}}\langle \exp[i\theta(t)]\rangle_{{\bm q}_{\rm in}}$, which in turn can be obtained from the result~\eqref{eq:result-cond_avg-Atheta} with $s_2=0$. We obtain 

\begin{align}
\label{eq:sub-ensemble-avg-state}
&{}_{{\bm q}_{\rm f}}\langle z(t)\rangle_{{\bm q}_{\rm in}} = e^{-t(1-t/T)/2\tau_{\rm m}} \nonumber \\
&\hspace{0.75cm}\times\frac{\sum_{n\in\mathbb{Z}}e^{-(\Delta\theta + 2\pi n)^2\tau_{\rm m}/2T}\cos\left(\theta_{\rm in} + (\Delta\theta + 2\pi n)t/T\right)}{\sum_{n\in\mathbb{Z}}e^{-(\Delta\theta + 2\pi n)^2\tau_{\rm m}/2T}},\nonumber \\
&{}_{{\bm q}_{\rm f}}\langle x(t)\rangle_{{\bm q}_{\rm in}} = e^{-t(1-t/T)/2\tau_{\rm m}} \nonumber \\
&\hspace{0.75cm}\times\frac{\sum_{n\in\mathbb{Z}}e^{-(\Delta\theta  + 2\pi n)^2\tau_{\rm m}/2T}\sin(\theta_{\rm in} + (\Delta\theta + 2\pi n)t/T)}{\sum_{n\in\mathbb{Z}}e^{-(\Delta\theta + 2\pi n)^2\tau_{\rm m}/2T}}, 
\end{align}
and ${}_{{\bm q}_{\rm f}}\langle y(t)\rangle_{{\bm q}_{\rm in}}=0$ since we are considering state evolution on the Bloch $xz$ plane. 

In Fig.~\ref{fig-1} we show some of the results obtained in this section. Figure~\ref{fig-1}(a) depicts the sub-ensemble average state for various values of $T$ with boundary conditions on the quantum trajectories specified by the angles $\theta_{\rm in}=\pi/4$ and $\theta_{\rm f}=7\pi/8$. We notice that for $T = 10\tau_{\rm m}$, the average ${\bm q}_{\rm sub-ens-avg}(t)$ first becomes mixed (i.e., states lying inside the Bloch sphere) and approaches the fully mixed state ($x=y=z=0$) at around $t = T/2$, and then it unwinds itself such that the subensemble average state reaches the target pure state ${\bm q}_{\rm f}$ at the final time $T$. This turning behavior is less obvious for shorter post-selection time $T = 3.5 \tau_{\rm m}$ and $T = \tau_{\rm m}$ as the trajectories in the subensemble do not have enough time to wander around the Bloch sphere to contribute to the mixedness of the state. We point out that this evolution is rather different from the conventional exponential decay of the (total) ensemble average evolution induced by the XZ measurement (shown as the dashed line in Figure~\ref{fig-1}(a)). Figure~\ref{fig-1}(b) shows the quantum state correlators for the ideal XZ measurement with the same boundary conditions as in Fig.~\ref{fig-1}(a) and $T=3.5\tau_{\rm m}$. Solid lines depict our analytical formulas, Eqs.~\eqref{eq:Czz-v2}--\eqref{eq:Cxz-v2} and~\eqref{eq:result-cond_avg-Atheta}, and the crosses depict Monte Carlo simulation results. The percentage of simulated trajectories which satisfy the boundary conditions are $0.12\%$, $0.30\%$ and $0.34\%$ for $T=\tau_{\rm m}, 3.5\tau_{\rm m}$ and $10\tau_{\rm m}$, respectively. 

\subsection{Perturbative solutions of conditional averages for non-ideal XZ measurements}\label{sec-nonideal}
\begin{table*}
{\renewcommand{\arraystretch}{1.8}
\begin{tabular}{  |>{\centering\arraybackslash} m{2cm} |>{\centering\arraybackslash} m{3.2cm} | >{\centering\arraybackslash} m{3.8cm}|  >{\centering\arraybackslash} m{2.5cm} | }
\hline
Type & Labels of vertices & Value & Diagrams  \\ \hline
Initial & $ x_{\rm in}, z_{\rm in}$ & $  x_{\rm in}, z_{\rm in} $ & 
\begin{tikzpicture}[node distance=0.6cm and 0.8cm]
\coordinate (b2);
\coordinate[right=0.5cm of b2] (bp);
\draw[particle] (b2) -- (bp);
\draw (bp) circle (.06cm);
\end{tikzpicture} \\ \hline
2 legged  & $p_{x} \xi_x, p_{z} \xi_z $ & $\frac{1}{\sqrt{\tau_x}}, \frac{1}{\sqrt{\tau_z}}$ & \begin{tikzpicture}[node distance=0.3cm and 0.5cm]
\coordinate (b2);
\coordinate[right=0.5cm of b2] (bp);
\coordinate[below=0.35cm of bp](bpp);
\coordinate[right=of bp](c0);
\draw[particle] (b2) -- (bp);
\draw[gluon](bp)--(bpp);
\draw (bp) circle (.06cm);
\end{tikzpicture} \\ \hline
4 legged & $p_z z x \xi_x$, $p_x x^2 \xi_x $,  $p_x x z \xi_z$, $p_z z^2 \xi_z$ & $-\frac{1}{\sqrt{\tau_x}},  -\frac{1}{\sqrt{\tau_z}}$ & 
\begin{tikzpicture}[node distance=0.3cm and 0.5cm]
\coordinate (b2);
\coordinate[right=0.5cm of b2] (bp);
\coordinate[below=0.5cm of bp](bpp);
\coordinate[right=of bp](c0);
\coordinate[below=0.2cm of c0](c1);
\coordinate[above=0.2cm of c0](c2);
\draw[particle] (b2) -- (bp);
\draw[gluon](bp)--(bpp);
\draw (bp) circle (.06cm);
\draw[particle2](bp) sin (c1);
\draw[particle2](bp) sin (c2);
\end{tikzpicture}  \\ \hline 
\end{tabular}}
\caption{Different possible vertices and associated diagrams for correlation functions of $x$ and $z$ in joint continuous XZ measurement. We note the different measurement times $\tau_{x,z}$ go with the appropriate $x$ or $z$ diagram vertex or propagator.}
\label{table1}
\end{table*}

In the case of non-ideal XZ measurements, where there exist measurement inefficiencies, qubit energy  relaxation and dephasing due to unwanted coupling with the environment, our knowledge about the qubit state comes in the form of a mixed state. In order to describe its evolution, we need to consider two coordinates; for example, the $x$ and $z$ Bloch coordinates. In this case, an analytic solution for the correlators is not forthcoming, so we apply the stochastic path integral perturbatively to compute the conditional averages for the quantum state variables. Following the method presented in Refs.~\cite{chantasri2015stochastic,jordan2015fluores}, the stochastic action discussed in Section~\ref{sec-spi}, Eqs.~\eqref{eq-actionXZ1}--\eqref{eq-actionXZ2}, is first rearranged into a free action and an interaction action, ${\cal S} = {\cal S}_F + {\cal S}_I$, where,
\begin{align}
\label{eq-freeaction} {\cal S}_F = \int_0^T\!\! dt & \left\{ i p_x ({\dot x}+\Gamma_z x) + i p_z({\dot z} + \Gamma_x z) + \frac{\xi_x^2}{2} + \frac{\xi_z^2}{2}\right\},\\
{\cal S}_I = \int_0^T \!\! dt &\left[\left\{ i p_x x z   - i p_z (1-z^2)  \right\} \frac{\xi_z}{\sqrt{\tau_z}} \right. \nonumber \\
&+ 
\left. \left\{-i p_x (1-x^2)    + i p_z x z \right\} \frac{\xi_x}{\sqrt{\tau_x}}\right].\label{eq-intaction}
\end{align}
The free action includes only the bilinear terms in $x,z,p_x,p_z,\xi_x$ and $ \xi_z$ variables and therefore can be rewritten in terms of the free Green's function, e.g., ${\cal S}_F = i \sum_{a = x,z} \int \!\! dt dt' p_a(t) G_a^{-1}(t,t') a(t')$, where $a = x,z$. The rest of the terms in the action define the interaction action ${\cal S}_I$. The free Green's functions for the variables $x$ and $z$ are given by
\begin{align} \label{eq-greenfn}
G_x(t, t') = \exp \left\{- \Gamma_z (t-t') \right\}\Theta(t - t'),\nonumber \\
G_z(t, t') = \exp \left\{- \Gamma_x (t-t') \right\}\Theta(t - t').
\end{align}
 The function $\Theta(t)$ is a left continuous Heaviside step function ($\Theta(0) =0$ and $\lim_{t\rightarrow 0^+} \Theta(t) = 1$, see Ref.~\cite{chantasri2015stochastic}). The Green's functions for the noise terms in the free action are simply the delta functions $G_{\xi}(t,t') = \delta(t-t')$, for both noises $\xi_x$ and $ \xi_z$. The perturbative expansion comes from expanding the exponential, $e^{{\cal S}_I}$, in powers of ${\cal S}_I$, where one can construct diagrammatic rules to keep track of nonvanishing terms. In the diagrammatic expansion, the terms in the interaction action ${\cal S}_I$ determine vertices, whereas, the Green's functions determine propagators. The vertices and propagators are shown in Table~\ref{table1}. We follow the diagrammatic rules explained in full detail in Ref.~\cite{chantasri2015stochastic} and note that the type of expansion presented in the reference is similar to the loop expansion in quantum theory. However, one can show that, in this case, the order of the expansion can be equivalently controlled by a small noise parameter which is related to the measurement efficiency $\eta_{x,z}$. Even though the measurement efficiency is not shown explicitly in the equations we used so far, they are actually contained in the definition of the characteristic measurement time $\tau_{x,z} = 1/(2 \Gamma_{x,z} \eta_{x,z})$.

We illustrate the diagrammatic approach by computing the correlators of this type: $C_{AB}(t_1,t_2|{\bm q}_{\rm in},0) \equiv \la A(t_1) B(t_2)\ra_{{\bm q}_{\rm in}} $, where $A, B$ are any two of the Bloch sphere coordinates $x,z$ and the conditioning is only on the initial state ${\bm q}_{\rm in} = \{ x_{\rm in}, z_{\rm in} \}$ at time $t_0=0$. We note that conditioning on both initial and final states is also possible, however the diagrammatic rules are far more complicated and we are not considering that case here. We begin with the cross-correlation function,
\begin{align}
C_{zx}(t_1,t_2|&{\bm q}_{\rm in},0) =  \, \la z(t_1) x(t_2) \ra_{{\bm q}_{\rm in}} = \begin{tikzpicture}[node distance=0.6cm and 0.8cm]
\coordinate[label=below:$x_{t_2}$] (b2);
\coordinate[right=of b2,label=below:$x_{\rm in}$] (bp);
\coordinate[above=of b2,label=above:$z_{t_1}$](a1);
\coordinate[above=of bp,label=above:$z_{\rm in}$](ap);
\draw[particle] (b2) -- (bp);
\draw[particle] (a1) -- (ap);
\draw (bp) circle (.06cm);
\draw (ap) circle (.06cm);
\fill[black] (b2) circle (.05cm);
\fill[black] (a1) circle (.05cm);
\end{tikzpicture} \nonumber \\
 +& 
\!\!\begin{tikzpicture}[node distance=0.6cm and 0.8cm]
\coordinate[label=below:$x_{t_2}$] (b2);
\coordinate[right=of b2] (bp);
\coordinate[above=of bp,label=below left:$z$] (ap);
\coordinate[left=1.1cm of ap,label=above:$z_{t_1}$] (a1);
\coordinate[right=of ap](cc);
\coordinate[right=of bp](bb);
\coordinate[below=0.1cm of cc](dp);
\coordinate[above=0.1cm of bb,label=right:$x_{\rm in}$](cpp);
\coordinate[below=0.3cm of bb,label=right:$z_{\rm in}$](dpp);
\draw[particle] (a1) -- (ap);
\draw[particle] (b2) -- (bp);
\draw[gluon] (ap) --  (bp);
\draw[particle2] (bp) sin  (dpp);
\draw[particle] (bp) sin  (cpp);
\fill[black] (a1) circle (.05cm);
\fill[black] (b2) circle (.05cm);
\draw (cpp) circle (.06cm);
\draw (dpp) circle (.06cm);
\end{tikzpicture}+ \begin{tikzpicture}[node distance=0.6cm and 0.8cm]
\coordinate[label=below:$x_{t_2}$] (b2);
\coordinate[right=of b2] (bp);
\coordinate[above=of bp,label=below left:$x$] (ap);
\coordinate[left=1.1cm of ap,label=above:$z_{t_1}$] (a1);
\coordinate[right=of ap](cc);
\coordinate[right=of bp](bb);
\coordinate[above=0.3cm of cc,label=right:$x_{\rm in}$](cp);
\coordinate[below=0.1cm of cc,label=right:$z_{\rm in}$](dp);
\draw[particle] (a1) -- (ap);
\draw[particle] (b2) -- (bp);
\draw[gluon] (ap) --  (bp);
\draw[particle2] (ap) sin  (dp);
\draw[particle] (ap) sin  (cp);
\fill[black] (a1) circle (.05cm);
\fill[black] (b2) circle (.05cm);
\draw (cp) circle (.06cm);
\draw (dp) circle (.06cm);
\end{tikzpicture}\nonumber\\
+& \begin{tikzpicture}[node distance=0.6cm and 0.8cm]
\coordinate[label=below:$x_{t_2}$] (b2);
\coordinate[right=of b2] (bp);
\coordinate[above=of bp,label=below left:$z$] (ap);
\coordinate[left=1.1cm of ap,label=above:$z_{t_1}$] (a1);
\coordinate[right=of ap](cc);
\coordinate[right=of bp](bb);
\coordinate[above=0.3cm of cc,label=right:$z_{\rm in}$](cp);
\coordinate[below=0.1cm of cc,label=right:$z_{\rm in}$](dp);
\coordinate[above=0.1cm of bb,label=right:$x_{\rm in}$](cpp);
\coordinate[below=0.3cm of bb,label=right:$z_{\rm in}$](dpp);
\draw[particle] (a1) -- (ap);
\draw[particle] (b2) -- (bp);
\draw[gluon] (ap) --  (bp);
\draw[particle2] (ap) sin  (dp);
\draw[particle] (ap) sin  (cp);
\draw[particle2] (bp) sin  (dpp);
\draw[particle] (bp) sin  (cpp);
\fill[black] (a1) circle (.05cm);
\fill[black] (b2) circle (.05cm);
\draw (cp) circle (.06cm);
\draw (dp) circle (.06cm);
\draw (cpp) circle (.06cm);
\draw (dpp) circle (.06cm);
\end{tikzpicture} + \begin{tikzpicture}[node distance=0.6cm and 0.8cm]
\coordinate[label=below:$x_{t_2}$] (b2);
\coordinate[right=of b2] (bp);
\coordinate[above=of bp,label=below left:$x$] (ap);
\coordinate[left=1.1cm of ap,label=above:$z_{t_1}$] (a1);
\coordinate[right=of ap](cc);
\coordinate[right=of bp](bb);
\coordinate[above=0.3cm of cc,label=right:$x_{\rm in}$](cp);
\coordinate[below=0.1cm of cc,label=right:$z_{\rm in}$](dp);
\coordinate[above=0.1cm of bb,label=right:$x_{\rm in}$](cpp);
\coordinate[below=0.3cm of bb,label=right:$x_{\rm in}$](dpp);
\draw[particle] (a1) -- (ap);
\draw[particle] (b2) -- (bp);
\draw[gluon] (ap) --  (bp);
\draw[particle2] (ap) sin  (dp);
\draw[particle] (ap) sin  (cp);
\draw[particle2] (bp) sin  (dpp);
\draw[particle] (bp) sin  (cpp);
\fill[black] (a1) circle (.05cm);
\fill[black] (b2) circle (.05cm);
\draw (cp) circle (.06cm);
\draw (dp) circle (.06cm);
\draw (cpp) circle (.06cm);
\draw (dpp) circle (.06cm);
\end{tikzpicture}, 
\end{align}
showing how the ending vertices $z_{t_1}$ and $x_{t_2}$ can be connected to other vertices in Table~\ref{table1} only up to $0$ loops (tree-level diagrams). Importantly, we note that the noise propagators (shown as wavy lines) can only connect vertices with the same noise flavors, as a result of the independence of the two noises $\xi_x$ and $\xi_z$.  The mathematical expressions corresponding to the diagrams are given by
\begin{align}
& C_{zx}(t_1,t_2|{\bm q}_{\rm in},0) =  -\left(\frac{1}{\tau_z}+\frac{1}{\tau_x}\right) z_{\rm in} x_{\rm in}\nonumber \\
&\times \int_0^T\!\! dt' G_z(t_1,t') G_x(t_2, t') G_x(t',t_0) G_z(t',t_0) \nonumber \\
&+ \frac{ x_{\rm in} z_{\rm in}^3}{\tau_z} \int_0^T \!\!dt' G_z(t_1,t') G_x(t_2, t') G_z(t',t_0)^3 G_x(t',t_0) \nonumber \\
&+ \frac{x_{\rm in}^3 z_{\rm in}}{\tau_x}  \int_0^T \!\! dt' G_z(t_1,t') G_x(t_2, t') G_z(t',t_0) G_x(t',t_0)^3. 
\end{align}
We evaluate above integrals and obtain
\begin{align}
&{C}_{zx} =  e^{-\Gamma_x t_1 - \Gamma_z t_2}\big\{-2 x_{\rm in} z_{\rm in} \left(\Gamma_z \eta_z + \Gamma_x \eta_x \right)
t_{\rm min} + \nonumber\\
& \frac{x_{\rm in} z_{\rm in}^3 \Gamma_z \eta_z}{ \Gamma_x} \left(1- e^{-2 \Gamma_x t_{\rm min}}\right) 
+ \frac{z_{\rm in} x_{\rm in}^3 \Gamma_x \eta_x}{ \Gamma_z} \left(1 - e^{-2 \Gamma_z t_{\rm min}}\right)
\big\}.
\end{align}
Other examples are the self-correlators $C_{zz}$ and $C_{xx}$. The diagrams for the $z$-$z$ correlator conditioning on an initial state are given by
\begin{align}
C_{zz}(t_1,t_2|{\bm q}_{\rm in}&,0)  =  \, \la z(t_1) z(t_2) \ra_{{\bm q}_{\rm in}} = \begin{tikzpicture}[node distance=0.6cm and 0.8cm]
\coordinate[label=below:$z_{t_2}$] (b2);
\coordinate[right=of b2,label=below:$z_{\rm in}$] (bp);
\coordinate[above=of b2,label=above:$z_{t_1}$](a1);
\coordinate[above=of bp,label=above:$z_{\rm in}$](ap);
\draw[particle] (b2) -- (bp);
\draw[particle] (a1) -- (ap);
\draw (bp) circle (.06cm);
\draw (ap) circle (.06cm);
\fill[black] (b2) circle (.05cm);
\fill[black] (a1) circle (.05cm);
\end{tikzpicture} + 
\!\!\begin{tikzpicture}[node distance=0.6cm and 0.8cm]
\coordinate[label=below:$z_{t_2}$] (b2);
\coordinate[right=of b2] (bp);
\coordinate[above=of bp,label=below left:$z$] (ap);
\coordinate[left=1.1cm of ap,label=above:$z_{t_1}$] (a1);
\coordinate[right=of ap](cc);
\coordinate[right=of bp](bb);
\coordinate[below=0.1cm of cc](dp);
\coordinate[above=0.1cm of bb](cpp);
\coordinate[below=0.3cm of bb](dpp);
\draw[particle] (a1) -- (ap);
\draw[particle] (b2) -- (bp);
\draw[gluon] (ap) --  (bp);
\fill[black] (a1) circle (.05cm);
\fill[black] (b2) circle (.05cm);
\end{tikzpicture}\nonumber \\
\nonumber +& 
\!\!\begin{tikzpicture}[node distance=0.6cm and 0.8cm]
\coordinate[label=below:$z_{t_2}$] (b2);
\coordinate[right=of b2] (bp);
\coordinate[above=of bp,label=below left:$z$] (ap);
\coordinate[left=1.1cm of ap,label=above:$z_{t_1}$] (a1);
\coordinate[right=of ap](cc);
\coordinate[right=of bp](bb);
\coordinate[below=0.1cm of cc](dp);
\coordinate[above=0.1cm of bb,label=right:$z_{\rm in}$](cpp);
\coordinate[below=0.3cm of bb,label=right:$z_{\rm in}$](dpp);
\draw[particle] (a1) -- (ap);
\draw[particle] (b2) -- (bp);
\draw[gluon] (ap) --  (bp);
\draw[particle2] (bp) sin  (dpp);
\draw[particle] (bp) sin  (cpp);
\fill[black] (a1) circle (.05cm);
\fill[black] (b2) circle (.05cm);
\draw (cpp) circle (.06cm);
\draw (dpp) circle (.06cm);
\end{tikzpicture}+ \begin{tikzpicture}[node distance=0.6cm and 0.8cm]
\coordinate[label=below:$z_{t_2}$] (b2);
\coordinate[right=of b2] (bp);
\coordinate[above=of bp,label=below left:$z$] (ap);
\coordinate[left=1.1cm of ap,label=above:$z_{t_1}$] (a1);
\coordinate[right=of ap](cc);
\coordinate[right=of bp](bb);
\coordinate[above=0.3cm of cc,label=right:$z_{\rm in}$](cp);
\coordinate[below=0.1cm of cc,label=right:$z_{\rm in}$](dp);
\draw[particle] (a1) -- (ap);
\draw[particle] (b2) -- (bp);
\draw[gluon] (ap) --  (bp);
\draw[particle2] (ap) sin  (dp);
\draw[particle] (ap) sin  (cp);
\fill[black] (a1) circle (.05cm);
\fill[black] (b2) circle (.05cm);
\draw (cp) circle (.06cm);
\draw (dp) circle (.06cm);
\end{tikzpicture}\nonumber \\
+& \begin{tikzpicture}[node distance=0.6cm and 0.8cm]
\coordinate[label=below:$z_{t_2}$] (b2);
\coordinate[right=of b2] (bp);
\coordinate[above=of bp,label=below left:$z$] (ap);
\coordinate[left=1.1cm of ap,label=above:$z_{t_1}$] (a1);
\coordinate[right=of ap](cc);
\coordinate[right=of bp](bb);
\coordinate[above=0.3cm of cc,label=right:$z_{\rm in}$](cp);
\coordinate[below=0.1cm of cc,label=right:$z_{\rm in}$](dp);
\coordinate[above=0.1cm of bb,label=right:$z_{\rm in}$](cpp);
\coordinate[below=0.3cm of bb,label=right:$z_{\rm in}$](dpp);
\draw[particle] (a1) -- (ap);
\draw[particle] (b2) -- (bp);
\draw[gluon] (ap) --  (bp);
\draw[particle2] (ap) sin  (dp);
\draw[particle] (ap) sin  (cp);
\draw[particle2] (bp) sin  (dpp);
\draw[particle] (bp) sin  (cpp);
\fill[black] (a1) circle (.05cm);
\fill[black] (b2) circle (.05cm);
\draw (cp) circle (.06cm);
\draw (dp) circle (.06cm);
\draw (cpp) circle (.06cm);
\draw (dpp) circle (.06cm);
\end{tikzpicture} + \begin{tikzpicture}[node distance=0.6cm and 0.8cm]
\coordinate[label=below:$z_{t_2}$] (b2);
\coordinate[right=of b2] (bp);
\coordinate[above=of bp,label=below left:$x$] (ap);
\coordinate[left=1.1cm of ap,label=above:$z_{t_1}$] (a1);
\coordinate[right=of ap](cc);
\coordinate[right=of bp](bb);
\coordinate[above=0.3cm of cc,label=right:$x_{\rm in}$](cp);
\coordinate[below=0.1cm of cc,label=right:$z_{\rm in}$](dp);
\coordinate[above=0.1cm of bb,label=right:$x_{\rm in}$](cpp);
\coordinate[below=0.3cm of bb,label=right:$z_{\rm in}$](dpp);
\draw[particle] (a1) -- (ap);
\draw[particle] (b2) -- (bp);
\draw[gluon] (ap) --  (bp);
\draw[particle2] (ap) sin  (dp);
\draw[particle] (ap) sin  (cp);
\draw[particle2] (bp) sin  (dpp);
\draw[particle] (bp) sin  (cpp);
\fill[black] (a1) circle (.05cm);
\fill[black] (b2) circle (.05cm);
\draw (cp) circle (.06cm);
\draw (dp) circle (.06cm);
\draw (cpp) circle (.06cm);
\draw (dpp) circle (.06cm);
\end{tikzpicture}, 
\end{align}
which can be evaluated to give
\begin{align}
&C_{zz} = e^{-\Gamma_x (t_1+t_2)} \large\{  - 4 \Gamma_z \eta_z \frac{z_{\rm in}^2}{\tau_z}   t_{\rm min}  + \frac{\Gamma_z \eta_z}{ \Gamma_x} \left( e^{ 2 \Gamma_x t_{\rm min}}-1 \right)\nonumber\\
& + \frac{x_{\rm in}^2 z_{\rm in}^2 \Gamma_x \eta_x}{ \Gamma_z}  \left(1 -  e^{- 2 \Gamma_z t_{\rm min}}\right) 
+  \frac{ z_{\rm in}^4 \Gamma_z \eta_z}{ \Gamma_x }   \left(1 -  e^{- 2 \Gamma_x t_{\rm min}} \right)\large\}.  
\end{align}
For the other self-correlator $C_{xx}$, the calculation is in the same way, with the replacements $x_{\rm in} \leftrightarrow z_{\rm in}$, and $z \leftrightarrow x$.

\begin{figure}
\includegraphics[width=8.7cm]{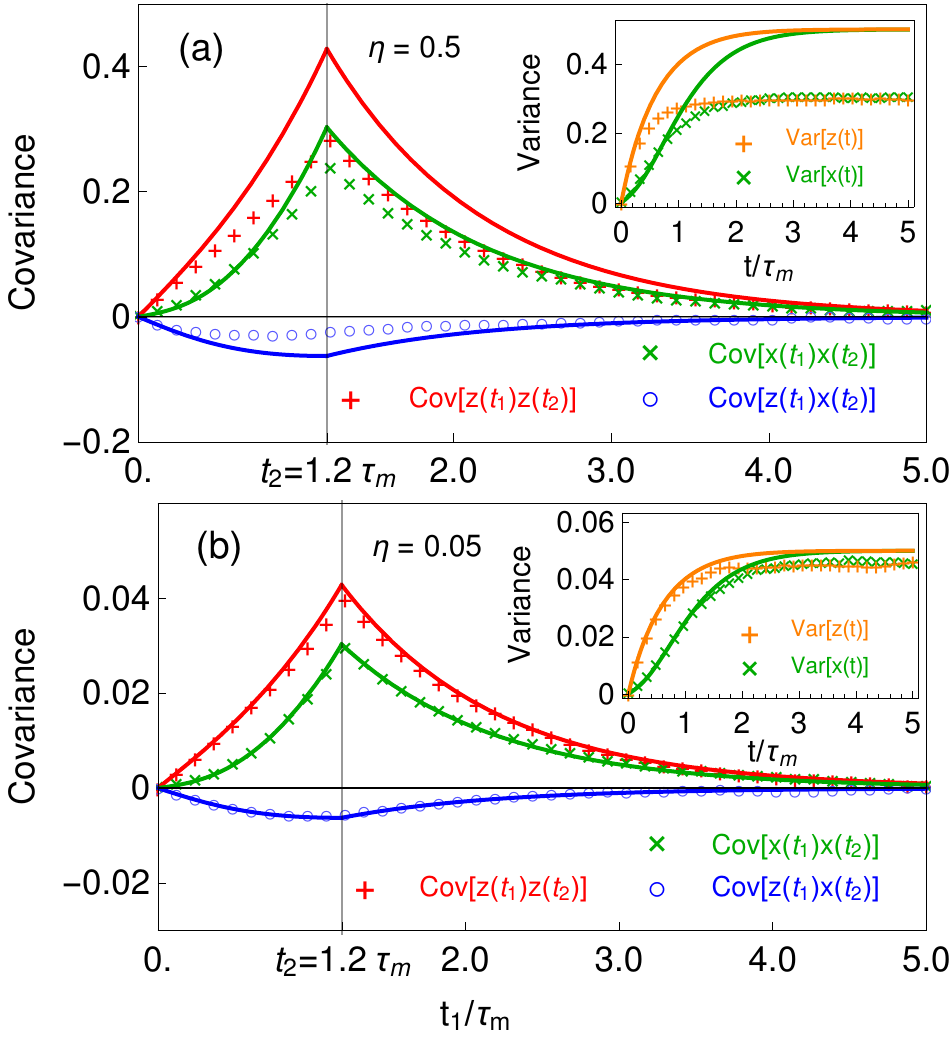}
\caption{Covariance and variance functions of the Bloch coordinates for non-ideal XZ measurements. The analytical results obtained from the diagrammatic perturbative expansion are presented in solid curves and the Monte Carlo simulation results are presented with markers. The results from the diagrammatic expansion agree with the numerical data much better for the case of small measurement quantum efficiency $\eta = 0.05$, as shown in panel (b), than for the case with $\eta = 0.5$, shown in panel (a). The insets shows the results for the variance functions. We assume both measurement channels have the same quantum efficiency ($\eta_x=\eta_z=\eta$). }
\label{fig-2}
\end{figure}

Figure~\ref{fig-2} shows the comparison between our analytical results from the diagrammatic perturbation theory and Monte Carlo simulations. We show the results for the covariance and variance functions of the Bloch coordinates, defined as ${\rm Cov}[A(t_1)B(t_2)] \equiv \la A(t_1) B(t_2) \ra_{{\bm q}_{\rm in}} - \la A(t_1) \ra_{{\bm q}_{\rm in}} \la B(t_2) \ra_{{\bm q}_{\rm in}}$ and ${\rm Var}[A(t)] \equiv \la \left[A(t) - \la A(t) \ra_{{\bm q}_{\rm in}}\right]^2\ra_{{\bm q}_{\rm in}}$, respectively. We consider two cases: measurement efficiencies $\eta_{x} = \eta_z = 0.5$ and 0.05. As expected from the expansion, where we only keep terms up to the tree-level diagrams, the numerical covariance and variance functions are in good agreement with the theory when $\eta_{x,z}$ are small, corresponding to the small noise limit where the noisy trajectories are not too far away from its averaged trajectory \cite{chantasri2015stochastic}.

\subsection{Application to general cases with non-commuting observables $\sigma_z$ and $\sigma_\varphi$}  \label{generalcase}
The analysis of the perturbative diagrams for XZ measurement can be generalized to the case of simultaneous and continuous measurement of $\sigma_z$ and $\sigma_\varphi$ with arbitrary angle, $\varphi$, between the measurement axes. The stochastic equations for this case are given in Eq.~\eqref{eq-itosde} (consider only equations for $x$ and $z$), and we can use them to construct a stochastic path integral with a stochastic action, which can be separated into free and interaction parts, $\tilde{\cal S} = \tilde{\cal S}_F + \tilde{\cal S}_I$. We notice that the terms in the free action for XZ measurement, Eq.~\eqref{eq-freeaction}, apart from the noise terms, are of the form: $ p_a ({\dot a} + \Gamma a)$ where $a = x,z$, leading to the Green's functions of Eq.~\eqref{eq-greenfn}. In order to be able to use the diagrammatic rules of XZ measurement case also in the case of arbitrary $\varphi$, we need to keep the same structure of the free action in both cases. We do this by transforming the variables $x,z$ in the stochastic master equations \eqref{eq-itosde} to a new set of variables $u,v$ that diagonalizes the linear parts of the evolution equations. That is, we find eigenvectors and eigenvalues of the decoherence matrix $M$,
\be
M = \begin{pmatrix} -( \Gamma_z + \Gamma_\varphi \cos^2 \varphi) & \frac{\Gamma_\varphi \sin 2 \varphi}{2} \\ \frac{\Gamma_\varphi \sin 2 \varphi}{2} & -\Gamma_\varphi \sin^2 \varphi
\end{pmatrix}.
\ee
The eigenvalues of the decoherence matrix are $\lambda_+$ and $\lambda_-$,
\begin{eqnarray}
\lambda_{\pm} &=& -\frac{\Gamma_z}{2} -\frac{\Gamma_\varphi}{2}  \pm \frac{\sqrt{\Xi}}{2}, 
\end{eqnarray}
where $\Xi = \Gamma_\varphi^2 + \Gamma_z^2 + 2 \Gamma_\varphi \Gamma_z \cos (2\varphi)$.
The stochastic differential equations after the variable transformation are of the form
\begin{subequations}\label{eq-odegeneraluv}
\begin{eqnarray}
{\dot u} &=& \lambda_- u + \kappa_{zu} \xi_z + \kappa_{xu} \xi_\varphi + \alpha_{x1} u^2 \xi_\varphi \nonumber \\
&+& \alpha_{x2} u v \xi_\varphi + \alpha_z u^2 \xi_z + \alpha_z u v \xi_z,\\
{\dot v} &=& \lambda_+ v + \kappa_{zv} \xi_z + \kappa_{xv} \xi_\varphi + \alpha_{x2} v^2 \xi_\varphi \nonumber \\
&+& \alpha_{x1} u v \xi_\varphi + \alpha_z v^2 \xi_z + \alpha_z u v \xi_z, 
\end{eqnarray}
\end{subequations}
where the $\alpha$- and $\kappa$-coefficients are functions of $\Gamma_z, \Gamma_\varphi$ and $\varphi$. The final form of the free action, constructed from these stochastic master equations, is given by $\tilde{\cal S}_F = \int_0^T \!\! dt \left\{i p_u ({\dot u} + \lambda_- u) +i  p_v ({\dot v} + \lambda_+ v ) + \xi_\varphi^2/2 + \xi_z^2/2\right\}$ which is similar to Eq.~\eqref{eq-freeaction}. Therefore, we can apply the same diagrammatic rules used in the XZ measurement case and only need to consider the new set of vertices built from the rest of the terms in the interaction action [those terms proportional to the $\alpha$- and $\kappa$-coefficients in Eq.~\eqref{eq-odegeneraluv}].

\section{Conditional averages using the Fokker-Planck Equation}
\label{fpe}
In this section we will use the Fokker-Planck equation for the transition probability, $P({\bm q},t|{\bm q}',t')$, to calculate the conditional quantum state correlators defined in Eqs.~\eqref{eq:Czz-v2}--\eqref{eq:Cxz-v2}. To find such correlators, we introduce the two-sided joint probability density $W_2$ for the quantum state to reach the states ${\bm q}_1$ at time $t_1$ and ${\bm q}_2$ at time $t_2\geq t_1$ given that the initial state is ${\bm q}_{\rm in}$ at time $t_0$ and the final state is ${\bm q}_{\rm f}$ at time $T$, 
\begin{align}
&W_2({\bm q}_{\rm f},T|{\bm q}_2,t_2;{\bm q}_1,t_1|{\bm q}_{\rm in},t_0)  =\nonumber \\
&\hspace{1cm} \frac{P({\bm q}_{\rm f},T|{{\bm q}_2,t_2})P({\bm q}_{\rm 2},t_2|{{\bm q}_1,t_1})P({\bm q}_{\rm 1},t_1|{{\bm q}_{\rm in},t_0})}{P({\bm q}_{\rm f},T|{{\bm q}_{\rm in},t_0})}.
\end{align}

The transition probability $P({\bm q},t|{{\bm q}',t'})$, with $t>t'$, can be obtained from the Fokker-Planck equation associated to the It\^o equations~\eqref{eq:EOM-x}--\eqref{eq:EOM-z}. In particular, such transition probability can be easily calculated analytically for the ideal XZ measurement case with detectors of equal measurement strengths. As in section~\ref{sec-corr}, we assume that the initial state is such that $y(0)=0$ and then $y(t)=0$ for all times. Quantum trajectories can then be parametrized by the polar coordinate, $\theta(t)$, which we will use it to denote the Bloch state ${\bm q} = \{\sin\theta,0,\cos\theta\}$. The Fokker-Planck equation associated to Eq.~\eqref{eq:theta-equ} reads as 
\begin{align}
\label{eq:FPE}
\partial_tP(\theta,t|\theta',t') = D\partial_\theta^2P(\theta,t|\theta',t'), 
\end{align}
where the diffusion coefficient is $D=(2\tau_{\rm m})^{-1}$ and the initial condition is $P(\theta,t'|\theta',t')=\delta(\theta-\theta')$. The solution of Eq.~\eqref{eq:FPE} is given by $P(\theta,t|\theta',t') = (2\pi)^{-1}\sum_{n\in \mathbb{Z}}\exp[in(\theta-\theta')-Dn^2(t-t')]$. The conditional average quantities, ${}_{{\bm q}_{\rm f}}\langle\mathcal{A}_{s_1s_2}[\theta(t)]\rangle_{{\bm q}_{\rm in}}$, in Eqs.~\eqref{eq:Czz-v2}--\eqref{eq:Cxz-v2} are now expressed in terms of the two-sided joint probability density as 
\begin{align}
\label{eq:As1s2-v2}
&{}_{{\bm q}_{\rm f}}\langle\mathcal{A}_{s_1s_2}[{\bm q}(\theta)]\rangle_{{\bm q}_{\rm in}} =\int_0^{2\pi}d\theta_1\int_0^{2\pi}d\theta_2\; e^{is_1\theta_1+is_2\theta_2}\times\nonumber \\ 
&\hspace{0.5cm}\frac{P(\theta_{\rm f},T|\theta_2,t_2)P(\theta_2,t_2|\theta_1,t_1)P(\theta_1,t_1|\theta_{\rm in},t_0)}{P(\theta_{\rm f},T|\theta_{\rm in},t_0)}.
\end{align}
It can be shown that Eq.~\eqref{eq:As1s2-v2} leads to the same result as Eq.~\eqref{eq:result-cond_avg-Atheta}. To show this, we evaluate the integrals in Eq.~\eqref{eq:As1s2-v2} and obtain (assuming $t_0=0$)
\begin{align}
\label{eq:derivation-1}
{}_{{\bm q}_{\rm f}}\langle\mathcal{A}_{s_1s_2}[{\bm q}(\theta)]\rangle_{{\bm q}_{\rm in}} =&\,e^{i\theta_{\rm in}(s_1+s_2) - D(s_1^2t_1 + s_2^2t_2 +2s_1s_2t_1)}\times \nonumber \\ 
&\frac{\sum_{n\in \mathbb{Z}}e^{-Dn^2T + [\Delta\theta - 2D(s_1t_1+s_2t_2)i]ni}}{\sum_{n\in\mathbb{Z}} e^{- Dn^2T+ \Delta \theta ni}}.
\end{align}
%
We then obtain the result~\eqref{eq:result-cond_avg-Atheta} after applying Poisson's resummation formula to the sums in the numerator and denominator in Eq.~\eqref{eq:derivation-1}.

\begin{figure}[tb!]
\centering
\includegraphics[width=0.9\linewidth, trim = 0cm 0cm 0cm 0cm,clip=true]{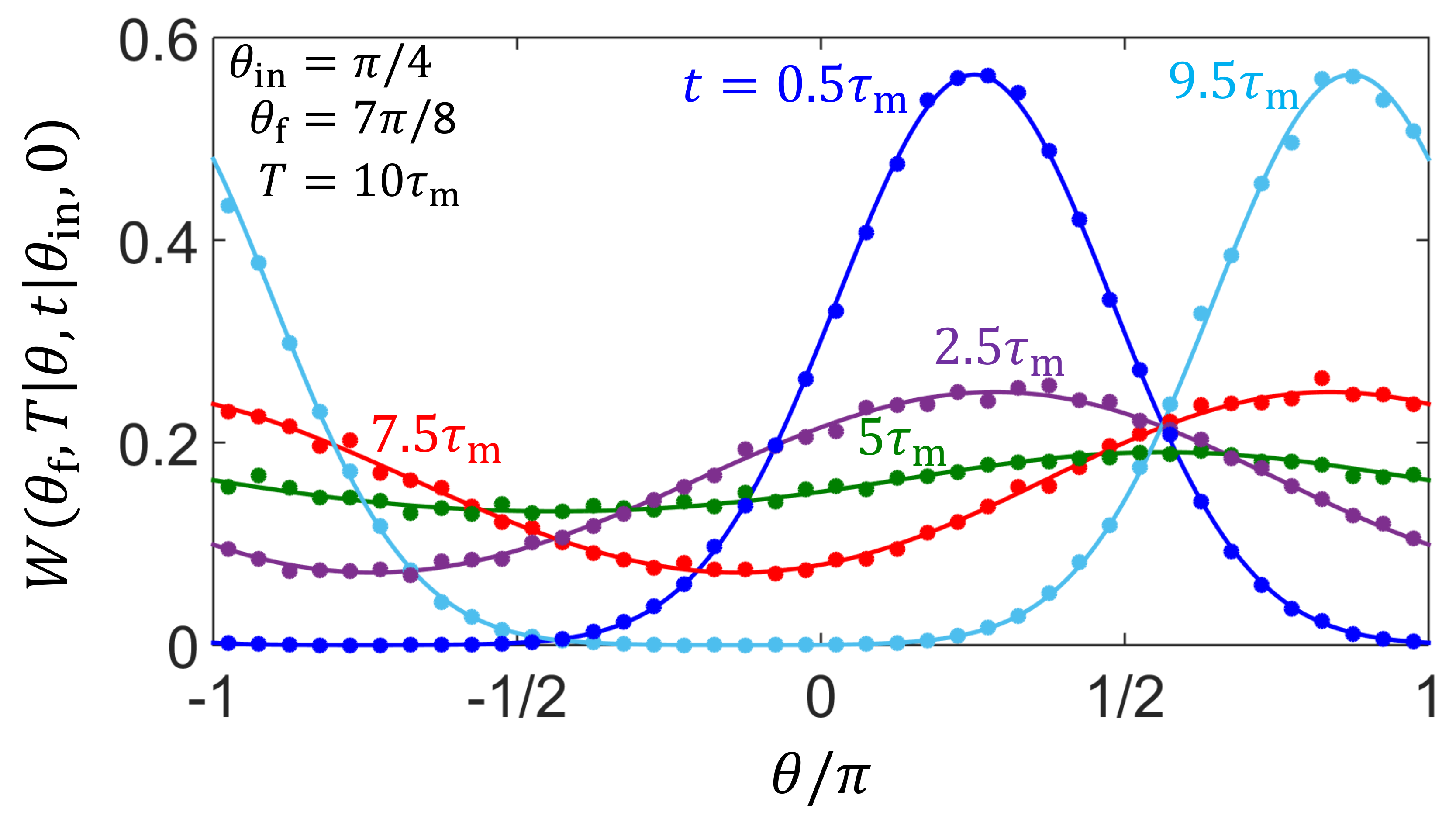}\\
\caption{Two-sided probability distribution of the polar coordinate, $\theta$, at the time $t$. Such distribution takes into account only quantum trajectories with fixed initial and final states, parametrized by the angles $\theta_{\rm  in}=\pi/4$ and $\theta_{\rm f}=7\pi/8$, at the times $t=0$ and $T=10\tau_{\rm m}$, respectively. The distribution is shown for various times: $t=0.5\tau_{\rm m}$, $2.5\tau_{\rm m}$, $5\tau_{\rm m}$, $7.5\tau_{\rm m}$ and $9.5\tau_{\rm m}$. The circles represent Monte Carlo simulation results. } 
\label{fig-3}
\end{figure}

The Fokker-Planck approach also enables us to find analytically the two-sided probability density $W$ of $\theta$ at the time $t$ in terms of the transition probability, obtained above,
\begin{align}
W({\theta}_{\rm f},T|\theta,t|\theta_{\rm in},t_0) = \frac{P(\theta_{\rm f},T|\theta,t)P(\theta,t|\theta_{\rm in},t_0)}{P(\theta_{\rm f},T|\theta_{\rm  in},t_0)}.
\end{align}
Figure~\ref{fig-3} shows the above distribution for various times between $t=0$ and $T=10\tau_{\rm m}$, and $\theta_{\rm in}=\pi/4$ and $\theta_{\rm f}=7\pi/8$. We note that, at short times, the considered distribution resembles a Gaussian-like distribution centered at the angle $\theta_{\rm in}$, then it tends to flatten out until the time $t=T/2$ (see green line in Fig.~\ref{fig-3}). For larger times, $t>T/2$, the distribution recovers a Gaussian-like shape centered at the angle $\theta_{\rm f}$. From Fig.~\ref{fig-3}, it is easy to understand the behavior exhibited by the sub-ensemble average state, ${\bm q}_{\rm sub-ens-avg}(t)$, shown in Fig.~\ref{fig-1}(a). Specifically, the decay of ${\bm q}_{\rm sub-ens-avg}$ toward the center of the Bloch sphere is due to the fact that the distribution, $W$, becomes increasingly flat until the time $t=T/2$. 

\section{Quantum state correlators for the transmon qubit experiment}\label{sec-exp}
\begin{figure}[t!]
\centering
\includegraphics[width=0.95\linewidth]{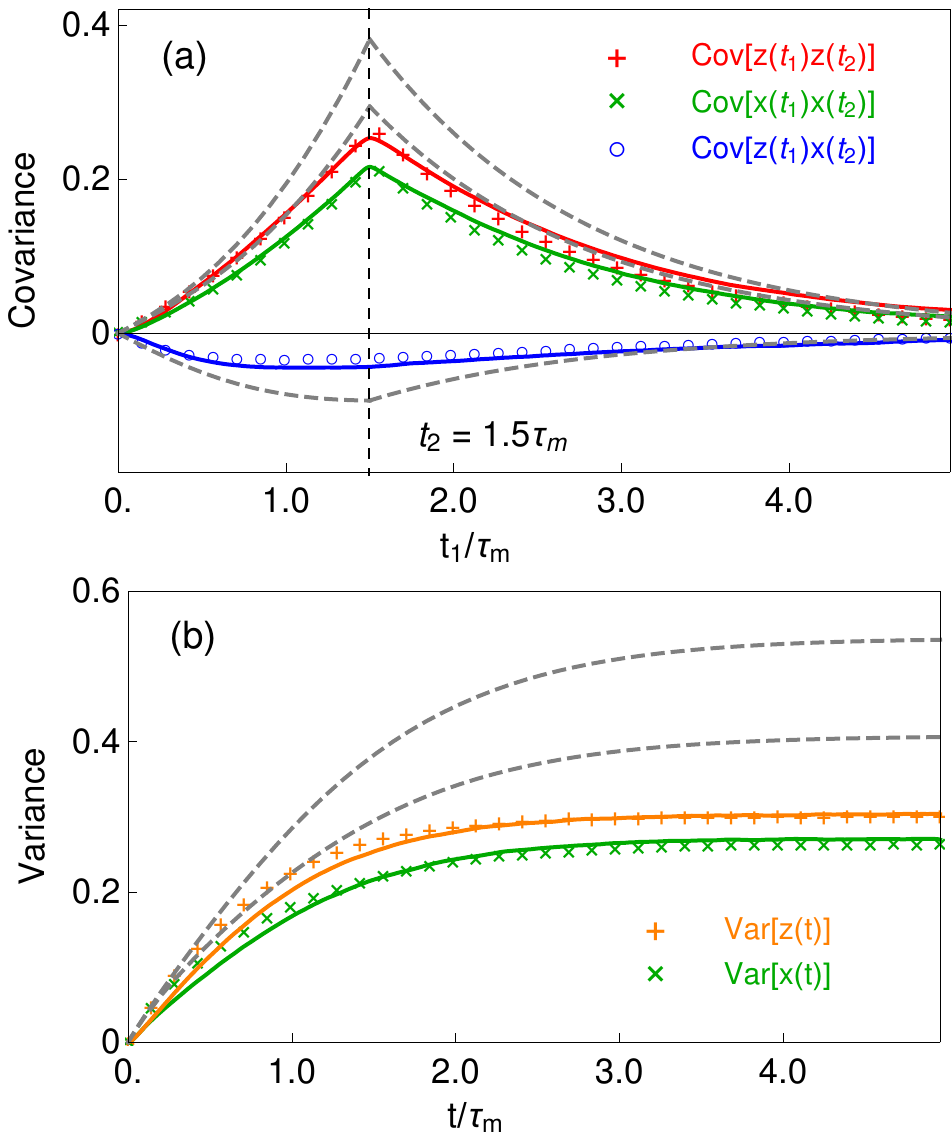}
\caption{Comparison of theory and experiment for quantum state correlators: (a) covariances and (b) variances, for the qubit under the simultaneous measurement of $\sigma_x$ and $\sigma_z$. The correlators derived from experimental qubit trajectories are shown in solid curves, whereas the Monte-Carlo simulation results are shown with markers, showing excellent agreement with the experimental data. As expected from Fig.~\ref{fig-2}, the first order perturbative solutions, shown in dashed gray, while giving the correct qualitative behaviour, deviate from the experimental results in the moderately high efficiency limit ($\eta_z=0.54$ and $\eta_x=0.41$). }
\label{fig:expt-vs-theory}
\end{figure}
We now discuss quantum state correlators for the experiment described in Ref.~\cite{Shay2016noncom}. In this experiment, simultaneous measurement of qubit observables $\sigma_z$ and $\sigma_x$ is implemented on a qubit rotating at the Rabi frequency, $\Omega_{\rm R}=2\pi\cdot 40$MHz. Both measurement axes also rotate on the $xz$ Bloch plane  at a frequency $\Omega_{\rm rf}$ close to $\Omega_{\rm R}$~\cite{Atalaya2017cor}. The measurement is a stroboscopic measurement where the measured qubit is an effective qubit living in the rotating frame defined by the frequency $\Omega_{\rm rf}$. As discussed in Ref.~\cite{Atalaya2017cor}, the effective qubit is subject to slow Rabi oscillations on the $xz$ plane because of the frequency mismatch: $\tilde\Omega_{\rm R} = \Omega_{\rm R} - \Omega_{\rm rf}\approx 2\pi\cdot 12$kHz, resulting in the effective qubit Hamiltonian $H_{\rm q}=\hbar\tilde\Omega_{\rm R}\sigma_y/2$. Moreover, the $T_1$- and $T_2$-relaxation processes of the physical qubit induce an effective depolarization channel on the effective qubit on the $xz$ plane, with a decay rate $\gamma = (T_1^{-1} + T_2^{-2})/2$ and $T_1=60\mu$s and $T_2=30\mu$s~\cite{Atalaya2017cor}. The residual Rabi oscillations and the $xz$-depolarization induces the qubit evolution as described by
\begin{subequations}
\begin{align}
\label{eq:residual-rabi-depolarization}
\dot x =& -\gamma x + \tilde\Omega_{\rm R}z \\
\dot z =& -\gamma z - \tilde\Omega_{\rm R}x.  
\end{align}
\end{subequations}
assuming no measurement occurred. Therefore, to include these effects to our theoretical model presented in Section~\ref{sec-sde}, the terms on the right side of the above equations can be simply added to the right side of Eqs.~\eqref{eq-itosde}.

However, since the experimental measurement readouts have finite integration time ($dt = 4$~ns), we cannot directly use the stochastic master equation to compute the quantum trajectory. Instead, the quantum trajectories are generated in discrete time-steps, i.e. a qubit state at the time $t+dt$ is obtained from a state at the time $t$ by means of the composed mapping: $\rho(t+dt) = \mathcal{L}_{\rm env}\circ\mathcal{M}[\rho(t)]$. Here, the operation $\mathcal{M}[\rho] = \mathcal{M}_x\circ\mathcal{M}_z[\rho]$ represents the evolution due to  measurement of $\sigma_z$ and $\sigma_x$ during the time period $dt$ (neglecting errors of order $dt^2$ due to the non-commutativity of the measured operators). The evolution due to measurement of $\sigma_z$ alone is defined as
\begin{align}
\label{eq:Bayesian-update}
\left(\mathcal{M}_z[\rho]\right)_{ij}= \frac{\left(M_z \rho M_z^{\dagger}\right)_{ij}}{{\rm Tr}[M_z\rho M_z^{\dagger}]}\exp(-\gamma_{ij}dt)
\end{align}
where $M_{z}=(4\pi\tau_z/dt)^{-1/2}\exp[-(r_{z}(t)-\sigma_z)^2dt/4\tau_z]$ and the matrix $\hat\gamma$ has vanishing diagonal elements ($\gamma_{00}=\gamma_{11}=0$) and off-diagonal elements equal to $\gamma_{01}=\gamma_{10}=\Gamma_z - 1/2\tau_z$. Note that $\hat\gamma$ vanishes in the case of ideal measurements. The state update Eq.~\eqref{eq:Bayesian-update} represents the quantum Bayesian update for a  non-ideal measurement of $\sigma_z$. The Bayesian update for $\sigma_x$-measurement is similar to the former one. First, we rotate the state along the $y$-axis by $-\pi/2$ [i.e., $\rho\to\exp(i\pi\sigma_y/4)\rho\exp(-i\pi\sigma_y/4)$], we apply the Bayesian update defined in Eq.~\eqref{eq:Bayesian-update}, and then we rotate back the state along the $y$-axis by $\pi/2$. For the current experiment, the measurement times are $\tau_z=1/(2\eta_z\Gamma_{z})=1.21\mu$s and $\tau_x = 1/(2\eta_x\Gamma_x)=1.60\mu$s (given that the measurement-induced dephasing rates are $\Gamma_z=\Gamma_x=1/1.3\mu$s). Another operation $\mathcal{L}_{\rm env}[\rho]$ represents the evolution due to environmental decoherence which in this case is obtained by evolving Eq.~\eqref{eq:residual-rabi-depolarization} over the time step, $dt$. In our case, the environment decoherence includes the effect of the quantum efficiencies $\eta_z=0.54,\eta_x=0.41$.  In the experiment, the initial Bloch vector is ${\bm q}_{\rm in}=\{\sin(\pi/4),0,\cos(\pi/4)\}$, and the total number of readout traces is approximately $2\times 10^5$ for each measurement channel. After calculating the quantum trajectories, we calculate the quantum correlators with only the pre-selection condition. The results are shown in Fig.~\ref{fig:expt-vs-theory}.

Figure~\ref{fig:expt-vs-theory}(a) shows the comparison between state correlators obtained from recorded readouts, Monte-Carlo simulations, and the perturbation theory discussed in section~\ref{sec-nonideal}. In the Monte-Carlo simulations, we use the same set of parameters used in constructing the qubit trajectories from the recorded readouts. We find excellent agreement between the simulation results and the results based on the recorded data. The perturbative results are still able to capture the correct trend of the correlators, cf. Fig.~\ref{fig-2}(a) even as a first order result. We note that by keeping higher order terms in the diagrammatic expansion will definitely improve the quantitative agreement. Fig.~\ref{fig:expt-vs-theory}(b) shows the variances of the Bloch coordinates $x$ and $z$. Here, we also find good agreement between the numerical simulations and the results obtained from the recorded readouts. 

\section{conclusions} \label{conc}
We have investigated the conditional averages and temporal correlation functions of the qubit state evolution under  simultaneous non-commuting observable measurement. 
We consider the joint measurement of the non-commuting pseudo-spin observables $\sigma_x$ and $\sigma_z$. Under these continuous measurements, the qubit state trajectories in time are described by the stochastic master equation, and the associated stochastic path integral.  In the ideal quantum limit measurement case of equal measurement rates, closed-form solutions of any multi-time correlation function can be found using the stochastic path integral formalism, even when conditioning on the initial and final states. We also presented a complimentary method for finding the conditional correlators using the Fokker-Planck equation. For the non-ideal measurement case, where the measurement is inefficient and environment dephasing is present, the conditional correlators can be obtained perturbatively using the diagrammatic approach. We have shown that the perturbative results, to first order in the efficiency, are in good agreement with the numerically simulated trajectories in the small quantum efficiency regime. Most importantly, we have compared the results of this theoretical analysis with the qubit trajectory data inferred and tomographically verified from a superconducting circuit coupled to multi-mode cavity, and have found excellent agreement between the correlation functions of the experimental data and our theoretical treatment.

\begin{acknowledgments}
This work was supported by US Army Research Office Grant No. W911NF-15-1-0496, by the National Science Foundation grant DMR-1506081, and the Templeton Foundation grant ID 58558. LM was supported by the National Science Foundation Graduate Fellowship Grant No. 1106400 and the Berkeley Fellowship for Graduate Study. 

*A.C. and J.A. contributed equally to this work.
\end{acknowledgments}


%

\end{document}